\documentclass[11pt]{article}

\usepackage{latexsym,amsfonts,amsmath,amssymb,enumerate,epsfig}
\usepackage{theorem}
\usepackage{mathrsfs}
\usepackage{ulem}
\usepackage{tikz}
\usepackage[all]{xy}

\usepackage{color}

\numberwithin{equation}{section}

\newcommand{\mb}[1]{\quad\mbox{#1}\quad}
\newcommand{\be}{\begin{equation}}
\newcommand{\ee}{\end{equation}}
\newcommand{\beu}{\begin{equation*}}
\newcommand{\eeu}{\end{equation*}}
\newcommand{\bea}{\begin{eqnarray}}
\newcommand{\eea}{\end{eqnarray}}
\newcommand{\beano}{\begin{eqnarray*}}
\newcommand{\eeano}{\end{eqnarray*}}
\newcommand{\bmx}{\begin{pmatrix}}
\newcommand{\emx}{\end{pmatrix}}

        \def\cC{{\cal C}}

\def\fC{{\mathfrak C}}

\newcommand{\wt}[1]{{\widetilde{#1}}}
\newcommand{\nonu}{\nonumber\\}
\newcommand{\ato}[2]{\genfrac{}{}{0pt}{}{#1}{#2}}
\newcommand{\va}{\mathbf{v}}
\newcommand{\wa}{\mathbf{u}}

\newcommand{\llangle}{\langle\!\langle}
\newcommand{\rrangle}{\rangle\!\rangle}
\newcommand{\steady}{|{\cal S}\rangle} 
\advance\textheight by \topskip
\begin{document}
 
\begin{center}

 {\LARGE  {\sffamily Inhomogeneous discrete-time exclusion processes} }\\[1cm]

\vspace{10mm}
  
{\Large 
 N. Crampe$^{a}$\footnote{nicolas.crampe@univ-montp2.fr}, 
 K. Mallick$^b$\footnote{kirone.mallick@cea.fr},
 E. Ragoucy$^{c}$\footnote{eric.ragoucy@lapth.cnrs.fr}
 and M. Vanicat$^{c}$\footnote{matthieu.vanicat@lapth.cnrs.fr}}\\[.41cm] 
{\large $^a$ Laboratoire Charles Coulomb (L2C), UMR 5221 CNRS-Universit\'e de Montpellier,\\[.242cm]
Montpellier, F-France.}
\\[.42cm]
{\large $^{b}$   Institut de Physique  Th\'eorique\\[.242cm] 
 CEA Saclay, F-91191 Gif-sur-Yvette, France. }
\\[.42cm]
{\large $^{c}$ Laboratoire de Physique Th{\'e}orique LAPTh
 CNRS and Universit{\'e} de Savoie.\\[.242cm]
   9 chemin de Bellevue, BP 110, F-74941  Annecy-le-Vieux Cedex, 
France. }
\end{center}
\vfill

\begin{abstract}
We study discrete time Markov processes with periodic or open boundary conditions and with inhomogeneous rates in the bulk.
The Markov matrices are given by the inhomogeneous transfer matrices introduced previously to prove the integrability of quantum spin chains.
We show that these processes have a simple graphical interpretation and correspond to a sequential update. We compute their stationary state
using a matrix ansatz and express their normalization factors as Schur polynomials. A connection between Bethe roots and Lee-Yang zeros is also 
pointed out.
\end{abstract}

\vfill\vfill
\rightline{LAPTh-033/15}
\rightline{June 2015}

\newpage
\pagestyle{plain}
\section*{Introduction} 
 
The study of systems far from equilibrium has greatly  benefited
from exact results obtained  for  interacting stochastic particle
systems \cite{GianniRevue,KLS,PaulK,Liggett,Zia,Spohn91} and amongst these,
the asymmetric simple exclusion process (ASEP) has become a
paradigm.  In one-dimension, the ASEP can be solved  exactly
and many physical quantities  (such as  its  phase diagram,
N-point correlations, density and current fluctuations etc...)
can be calculated analytically
\cite{CKZ,DerridaRep,DerrReview,DCairns,Schutz}.  Hence,  the
ASEP (as well as the closely related  Kardar-Parisi-Zhang
equation)  belongs to the   class of integrable models:  this
exceptional property explains its  solvability, its  intricate
algebraic  structure and  its  relations  to many different
fields  of  mathematics (representation theory, random matrices,
combinatorics...).  A new field, {\it  stochastic integrability},
at the crossroads of probability and algebra,  is indeed
emerging,   in which  the ASEP   plays   a  distinguished role \cite{Corwin}.

 Integrability is a fragile  property that is easily lost  when one
 tries to generalize or {\it deform} a model: a system has to respect
 stringent mathematical criteria in order to be integrable (basically,
 the Yang-Baxter relation in the bulk and reflection identities at the
 boundaries \cite{Faddeev,sklyanin}).  Our aim, in the present work,
 is to   study integrable, one  species exclusion processes with inhomogeneous and
non local transition rates. 
 We shall consider  both  cases of
 periodic (section \ref{sec:per}) and open boundary conditions
 (section \ref{sec:op}):  using the transfer matrix
 formalism, we  define  discrete time stochastic processes, by stating
 explicitly and  representing  graphically the  updating  rules  for
 the  dynamics.  Then, we  determine  the stationary states of these
 models exactly. When the inhomogeneities $z_i$ that enter 
 the transfer matrix are set to 1, the stationary measure coincides with the (periodic or open) TASEP one.

 For the open system, we construct  a matrix product
 Ansatz  \cite{DEHP,MartinRev}  using  the Zamolodchikov-Faddeev
 \cite{ZF} and the Ghoshal-Zamolodchikov  \cite{GZ} relations,  thanks to 
 the formalism developed in \cite{Sasamoto2,CRV,CMRV}.  Knowing the
 invariant measure, we  compute analytically some physical
 observables (such as the local density) that will be expressed in
 terms of Schur polynomials. Thus, our work  interrelates integrability,
 Matrix Ansatz  and  combinatorial properties of  interacting
 particle systems and  contributes to extend the class of exactly
 solvable models in non-equilibrium statistical mechanics. 
  We  also find  an unexpected relation between the Bethe roots (solutions
 of the Bethe equations found in \cite{crampe13}) and the Lee-Yang zeros
 of the normalization function  (see section \ref{sec:bly}).

\section{Markov process on the periodic lattice \label{sec:per}}

This section is devoted to the construction of a discrete time
 Markov process on the periodic lattice (the site $L+1$ is identified 
with the site $1$). After discussing the general framework  in section \ref{subsec:general}, 
 we  construct in  section \ref{subsec:Transfer} a one-parameter family of 
 commuting operators $t(x|\bar z)$  (using the integrability of the model).  
 Using these operators $t(x|\bar z)$, we define  a discrete time 
 Markov process  that can be interpreted  graphically (see {section} \ref{subsec:graph}).
 The stochastic updating rules corresponding 
to this process are given explicitly
 in {section} \ref{subsec:su}. 
Finally,  the stationary state of the process
  and some of its properties are studied
 in {section} \ref{subsec:stat}.

 \subsection{General framework: discrete time Markov processes}
 \label{subsec:general}

 We consider a one-dimensional lattice of size $L$. 
 Each site of the lattice can either be empty or carry one particle;
 a configuration of the system can be denoted by a $L$-tuple $\mathcal{C}=(\tau_1,\tau_2,\dots,\tau_L)$
 where we set $\tau_j=0$ if the site $j$ is empty and $\tau_j=1$ if it  is occupied. 
The configuration space is given by 
$\left(\mathbb{C}^2\right)^{\otimes L}=\underbrace{\mathbb{C}^2\otimes\dots\otimes\mathbb{C}^2}_{L \text{ times}}$ where the $j^{\text{th}}$ component
$\mathbb{C}^2$ of the tensor space represents the $j^{\text{th}}$ site of the lattice.
 A configuration of the system thus corresponds to a vector
 $|\tau_1\dots \tau_L\rangle=|\tau_1\rangle\otimes\dots\otimes|\tau_L\rangle$, where we have defined the basis vectors
$|0\rangle=\left(
 \begin{array}{c}
 1\\0
 \end{array}
\right)$ and $|1\rangle=\left(
 \begin{array}{c}
 0\\1
 \end{array}
\right)$.  


The dynamics  is stochastic: the system can evolve  from  a configuration $\mathcal{C}'$ to another configuration $\mathcal{C}$
with the probability $M(\mathcal{C},\mathcal{C}')$. The Markov property is satisfied:
 the probability $M(\mathcal{C},\mathcal{C}')$ depends only on the two
configurations $\mathcal{C}$ and $\mathcal{C}'$ and not on the previous history of the system.
Let  $P_t(\mathcal{C})$ be 
 the probability for the system to be in the
 configuration $\mathcal{C}$ at time $t$. Its (discrete)
 time evolution is given by 
\begin{equation}\label{eq:ME}
P_{t+1}(\mathcal{C})=\sum_{\mathcal{C}'\neq \mathcal{C}}M(\mathcal{C},\mathcal{C}')P_t(\mathcal{C}')
+\left(1 -\sum_{\mathcal{C}'\neq \mathcal{C}}M(\mathcal{C}',\mathcal{C})\right) P_t(\mathcal{C})
 = \sum_{\mathcal{C}'}M(\mathcal{C},\mathcal{C}')P_t(\mathcal{C}') \,,
\end{equation}
where
 the diagonal term $M(\mathcal{C},\mathcal{C})=1-\sum_{\mathcal{C}'\neq \mathcal{C}}M(\mathcal{C}',\mathcal{C})$,
 is   the probability of remaining in  configuration
 $\mathcal{C}$ between $t$ and $t+1$.
This equation can be rewritten in vector-form  as follows: defining 
\begin{equation}
 |P_t\rangle=\left(
 \begin{array}{c}
  P_t(\ (0,\dots,0,0,0)\ )\\
   P_t(\ (0,\dots,0,0,1)\ )\\
   P_t(\ (0,\dots,0,1,0)\ )\\
   \vdots\\
   P_t(\ (1,\dots,1,1,1)\ )
 \end{array}
\right)
=\sum_{\tau_1,\dots,\tau_L\in \{0,1\}}  P_t(\ (\tau_1,\dots,\tau_{L})\ )~ |\tau_1\dots \tau_L\rangle
\end{equation}
 the master equation \eqref{eq:ME} becomes 
\begin{equation}\label{eq:MEv}
 |P_{t+1}\rangle=M\ |P_t\rangle.
\end{equation}
 A fundamental quantity is  the stationary probability distribution of the process, i.e.,
  a vector $\steady$ such that $M\steady=\steady$
(the existence of this vector is guaranteed because the entries of each  column of $M$ sum to $1$).
 If the process is not reversible, that is if we do not have the 
detailed balance property $M(\mathcal{C}',\mathcal{C}){\cal S}(\mathcal{C}) \neq M(\mathcal{C},\mathcal{C}'){\cal S}(\mathcal{C}')$, the task of
finding an explicit expression for $\steady$ can be very hard. However in some  cases,
 in particular  when the Markov matrix is integrable, 
 this  stationary distribution can be calculated exactly.

\subsection{Inhomogeneous transfer matrix}
 \label{subsec:Transfer}
 
 We now use the theory of integrable systems to construct  a family of 
 operators,  $t(x|\bar z)$,  called  transfer matrices, that will allow us
 to define stochastic  updating rules. In this section, we discuss only about
 the integrability of the transfer matrix. Its Markov property will be discussed at the end of subsection \ref{subsec:graph}.

The building block of the transfer matrix is  the $R$-matrix: it acts on two sites and depends on a parameter $x$, called
 the spectral parameter. 
 Here,  we will consider the following $R$-matrix
 \begin{equation} \label{R-TASEP}
R(x)=\left( \begin {array}{cccc} 
1&0&0&0\\ 
0&0&x&0\\
0&1&1-x&0\\
0&0&0&1
\end {array} \right) .
\end{equation}
 It acts in the space $\mathbb{C}^2 \otimes \mathbb{C}^2$, where we choose the basis $|00\rangle$, $|01\rangle$,
 $|10\rangle$, $|11\rangle$ which respectively correspond to the following lattice configurations: $(0,0)$, $(0,1)$, $(1,0)$, $(1,1)$.
 The  $R$-matrix satisfies the Yang-Baxter equation:
 \begin{eqnarray}
 \label{eq:ybe}
 R_{12}\left(\frac{x_1}{x_2}\right)\ R_{13}\left(\frac{x_1}{x_3}\right)\
 R_{23}\left(\frac{x_2}{x_3}\right)\ =\ 
 R_{23}\left(\frac{x_2}{x_3}\right)\ R_{13}\left(\frac{x_1}{x_3}\right)
\ R_{12}\left(\frac{x_1}{x_2}\right)\ ,
\end{eqnarray}
where the subscripts denote on which tensor space the matrix $R$ has a non trivial action: for instance
$R_{12}=R\otimes 1$, $R_{23}=1\otimes R$ etc...
The $R$-matrix obeys the regularity relation:
 \begin{equation} \label{R-regularity}
 R(1)=P
\end{equation}
 where $P$ is the permutation operator ($P v\otimes w=w\otimes v$).
It also satisfies the unitarity relation:
\begin{equation}
 R_{12}\left(x\right)R_{21}\left(\frac{1}{x}\right)=1,
\end{equation}
where $R_{21}(x)=P_{12}R_{12}(x)P_{12}$.
The $R$-matrix \eqref{R-TASEP} is  usually associated with the 
 continuous time TASEP \cite{CRV}. 

From the  R-matrix, we  build an  inhomogeneous periodic 
  transfer matrix  as follows
\begin{equation}\label{eq:tp}
  t(x | \bar z)= tr_0 \left[ R_{0,L}\left(\frac{x}{z_L}\right)R_{0,L-1}\left(\frac{x}{z_{L-1}}\right)\dots R_{0,1}\left(\frac{x}{z_1}\right) \right].
 \end{equation}
The parameter $x$ is  the spectral parameter and the parameters $z_1,\dots, z_L$
 are called  inhomogeneities. 
 The homogeneous case  (\textit{i.e.} $z_i=1$)  was studied  in  \cite{GM1}. 
  
  The main feature of the transfer matrix is that, due to the Yang-Baxter equation
 \eqref{eq:ybe}, it commutes for different values of the spectral parameter.
This  is the key to integrability \cite{Faddeev}:
  $$[t(x|\bar z),t(y|\bar z)]=0 \, .$$ 
We now  introduce the following operator
\begin{eqnarray}
 M(x|\bar z) &=& t(x|\bar z) t(z_1|\bar z)^{-1}  \nonumber \\
&=&  tr_0 \left[ R_{0,L}\left(\frac{x}{z_L}\right)R_{0,L-1}\left(\frac{x}{z_{L-1}}\right)
\dots R_{0,1}\left(\frac{x}{z_1}\right) \right]
  R_{2,1}(\frac{z_2}{z_1})\dots R_{L,1}(\frac{z_L}{z_1}). \label{eq:M}
\end{eqnarray}
Note that we have normalized $t(x|\bar z) $ using $t(z_1|\bar z)$, but a different choice $t(z_j|\bar z)$, $j=2,3,...,L$ leads to a 
similar rotated matrix $M(x|\bar z)$.  
Obviously $M(x|\bar z)$  commutes with $t(y|\bar z)$ and
  has the same eigenvectors. We choose below to use the ``normalized matrix'' $M(x|\bar z)$ instead of $t(x|\bar z)$ 
because it allows one to construct easily local Hamiltonians (see e.g. remark below).

\paragraph{Remark:}
The operator $M(x|\bar z)$ is related
  to the continuous time  TASEP. Taking  the 
homogeneous limit $z_i \rightarrow 1$, we are left with $M(x)=t(x)t(1)^{-1}$, where
$t(x)=tr_0(R_{0,L}(x)\dots R_{0,1}(x))$. By standard computation,  we obtain 
\begin{equation}
 M'(1)=\sum_{k=1}^{L} P_{k,k+1}.R_{k,k+1}'(1),
\end{equation}
with the convention $L+1\equiv 1$. The operator
\begin{equation}
 -P.R'(1)=\left(\begin{array}{cccc}
           0&0&0&0\\
           0&0&1&0\\
           0&0&-1&0\\
           0&0&0&0
          \end{array}\right)
\end{equation}
is the local jump operator of the continuous time TASEP. Hence $-M'(1)$ is the Markov
 matrix of the continuous time TASEP. 
Therefore, all the results obtained below for the discrete time process
 will be  valid, when  $z_i\to 1$, for the continuous time TASEP.

\subsection{Graphical representation \label{subsec:graph}}

   The $R$-matrix and the transfer matrices defined above have 
a useful  graphical illustration  that we now explain.

\begin{figure}[htb]
\begin{center}
 \begin{tikzpicture}[scale=0.8]
\node at (-4,2) [thick] {$\langle00|R_{ij}(\frac{x}{y})|00\rangle$};
\node at (0,2) [thick] {$\langle10|R_{ij}(\frac{x}{y})|10\rangle$};
\node at (4,2) [thick] {$\langle01|R_{ij}(\frac{x}{y})|10\rangle$};
\node at (8,2) [thick] {$\langle10|R_{ij}(\frac{x}{y})|01\rangle$};
\node at (12,2) [thick] {$\langle11|R_{ij}(\frac{x}{y})|11\rangle$};
\foreach \i in {-4,0,8}
{\draw[dashed] (\i,0) -- (\i,1) ;}
\foreach \i in {-4,4}
{\draw[dashed] (\i,0) -- (\i+1,0) ;}
\foreach \i in {4,12}
{\draw[ultra thick] (\i,0) -- (\i,1) ;}
\foreach \i in {0,8,12}
{\draw[ultra thick] (\i,0) -- (\i+1,0) ;}
\foreach \i in {-4,0,4}
{\draw[->,dashed] (\i,-1) -- (\i,-0.5) ; \draw[dashed] (\i,-0.5) -- (\i,0) ;}
\foreach \i in {-4,8}
{\draw[->,dashed] (\i-1,0) -- (\i-0.5,0) ; \draw[dashed] (\i-0.5,0) -- (\i,0) ;}
\foreach \i in {8,12}
{\draw[->, ultra thick] (\i,-1) -- (\i,-0.5) ; \draw[ultra thick] (\i,-0.5) -- (\i,0) ;}
\foreach \i in {0,4,12}
{\draw[->, ultra thick] (\i-1,0) -- (\i-0.5,0) ; \draw[ultra thick] (\i-0.5,0) -- (\i,0) ;}
\foreach \i in {-4,0,4,8,12}
{\node at (\i-1.25,0) [] {\footnotesize{$i$}};\node at (\i,-1.25) [] {\footnotesize{$j$}};
\node at (\i-0.5,0.3) [] {\footnotesize{$x$}};\node at (\i +0.3,-0.5) [] {\footnotesize{$y$}};}
\node at (-4,-2.5) [thick] {\Large{$1$}};
\node at (0,-2.5) [thick] {\Large{$1-\frac{x}{y}$}};
\node at (4,-2.5) [thick] {\Large{$\frac{x}{y}$}};
\node at (8,-2.5) [thick] {\Large{$1$}};
\node at (12,-2.5) [thick] {\Large{$1$}};
 \end{tikzpicture}
 \end{center}
\caption{Non-vanishing vertices associated with  $R$. \label{matrix_elements}}
\end{figure}
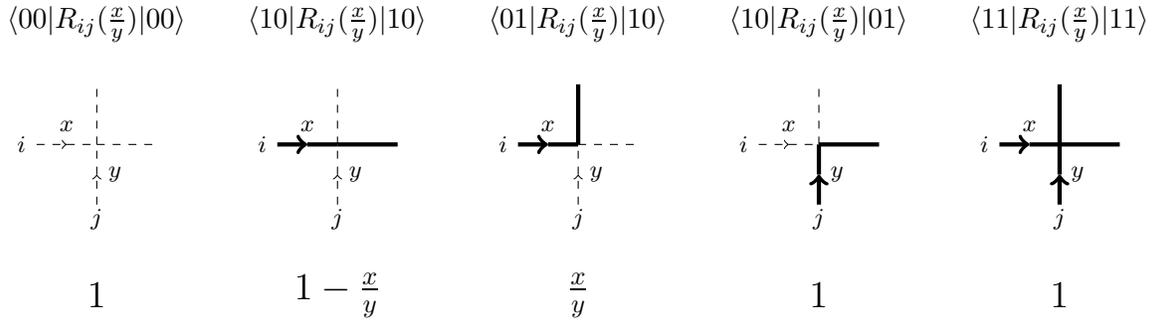

\paragraph{R-matrix.}
The action of the $R$-matrix $R_{ij}(x/y)$  can be represented
 by  the vertices given in  fig. \ref{matrix_elements}.
Each vertex corresponds to  a matrix element of the $R$-matrix. 
A dashed incoming line (according to the arrows direction) denotes the vector $|0\rangle$ (or equivalently an empty site),
whereas a continuous thick line denotes the vector $|1\rangle$ (or equivalently an occupied site).
In a similar way, the out-going lines (after the crossing point)
 represent the states $\langle0|$ and $\langle1|$ on which we are contracting 
 the matrix $R$.
The missing vertices in fig. \ref{matrix_elements} correspond
 to vanishing matrix elements of $R$.
Remark that the number of particles is conserved: in every non vanishing vertex, the number of incoming continuous thick lines is equal to
the number of out-going continuous thick lines.

With this graphical interpretation, we will be able to
 compute efficiently a matrix element
 of a product of $R$ matrices acting in different
components  of the tensor product. 

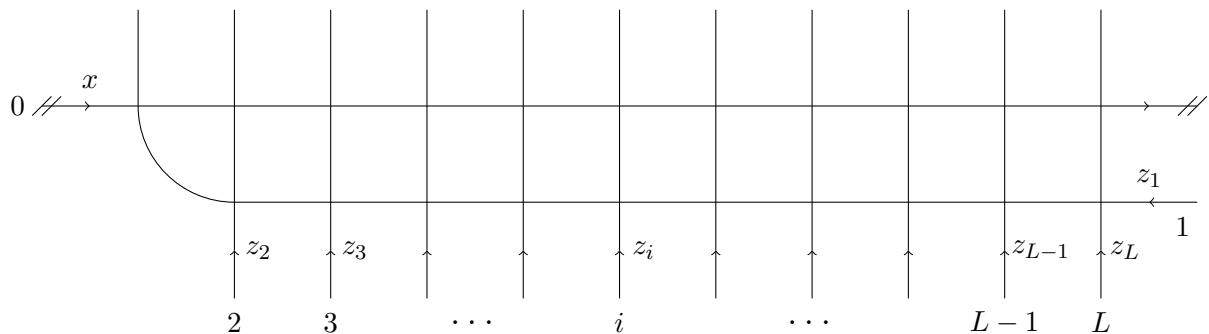
\begin{figure}[htb]
\begin{center}
 \begin{tikzpicture}[scale=0.64]
 \draw[] (-4,0) -- (16,0) ; 
 \draw[] (-2,-2) -- (16,-2) ;
 \draw[->] (-6,0) -- (-5,0) ; \draw[] (-5,0) -- (-4,0) ; 
 \draw[->] (16,0) -- (17,0) ; \draw[] (17,0) -- (18,0) ;
 \draw (17.8,-0.2) -- (18.2,0.2); \draw (17.6,-0.2) -- (18.0,0.2);
 \draw (-6.2,-0.2) -- (-5.8,0.2); \draw (-6,-0.2) -- (-5.6,0.2);
 \draw[] (-4,0) -- (-4,2) ; \draw[->] (18,-2) -- (17,-2) ;\draw[] (17,-2) -- (16,-2) ;
 \draw (-4,0) arc (180:270:2cm);
 \foreach \i in {-2,0,...,16}
{\draw[] (\i,0) -- (\i,2) ; \draw[->] (\i,-4) -- (\i,-3) ; \draw[] (\i,-3) -- (\i,0) ;}
\node at (-6.5,0) [thick] {$0$};
\node at (17.7,-2.5) [thick] {$1$};
\node at (-2,-4.5) [thick] {$2$};
\node at (0,-4.5) [thick] {$3$};
\node at (6,-4.5) [thick] {$i$};
\node at (3,-4.5) [thick] {\Large{$\dots$}};
\node at (10,-4.5) [thick] {\Large{$\dots$}};
\node at (14,-4.5) [thick] {$L-1$};
\node at (16,-4.5) [thick] {$L$};
\node at (-5,0.5) [thick] {$x$};
\node at (17,-1.5) [thick] {$z_1$};
\node at (-2+0.5,-3) [thick] {$z_2$};
\node at (0.5,-3) [thick] {$z_3$};
\node at (6.5,-3) [thick] {$z_i$};
\node at (14.75,-3) [thick] {$z_{L-1}$};
\node at (16.5,-3) [thick] {$z_L$};
 \end{tikzpicture}
 \end{center}
\caption{Graphical representation for $M(x|\bar z)$. Since the lattice is periodic, the two double slash should be considered
as linked.} \label{fig:rdm}
\end{figure}

\paragraph{Transfer matrix.}
From  expression \eqref{eq:M} we can deduce a graphical
representation for $M(x|\bar z)$. The starting point is the lattice
illustrated in fig. \ref{fig:rdm} that one has to fill according to
the matrix element one wants to compute. Instead of explaining it in
full generality, we take below a concrete example.

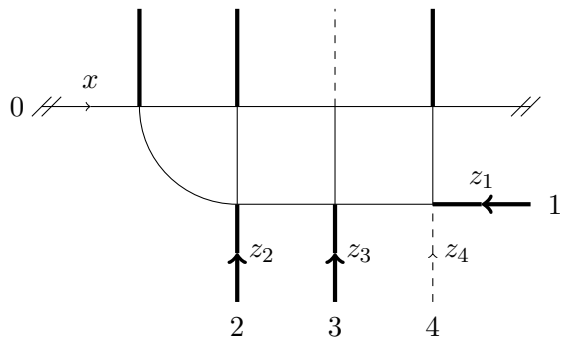
\begin{figure}[htb]
\begin{center}
 \begin{tikzpicture}[scale=0.65]
 \draw[->] (0,0) -- (1,0) ; 
 \draw[] (1,0) -- (10,0) ;
 \draw[ultra thick] (2,0) -- (2,2) ;
 \draw[->,ultra thick] (4,-4) -- (4,-3) ; \draw[ultra thick] (4,-3) -- (4,-2) ;\draw[] (4,-2) -- (4,0) ;\draw[ultra thick] (4,0) -- (4,2) ;
 \draw[->,ultra thick] (6,-4) -- (6,-3) ; \draw[ultra thick] (6,-3) -- (6,-2) ;
 \draw[] (8,-2) -- (8,0) ; \draw[ultra thick] (8,0) -- (8,2) ; 
 \draw[] (4,-2) -- (6,-2) ;
 \draw[->,ultra thick] (10,-2) -- (9,-2) ; \draw[ultra thick] (9,-2) -- (8,-2) ;
 \draw[] (2,0) arc (180:270:2cm);
 \draw[] (6,-2) -- (6,0); \draw[dashed] (6,0) -- (6,2);
 \draw[->,dashed] (8,-4) -- (8,-3);\draw[dashed] (8,-3) -- (8,-2);
 \draw[] (6,-2) -- (8,-2);
 \draw (9.8,-0.2) --(10.2,0.2);\draw (9.6,-0.2) --(10,0.2);
 \draw (-0.2,-0.2) --(0.2,0.2);\draw (0,-0.2) --(0.4,0.2);
 \node at (4.5,-3) [thick] {$z_2$};
 \node at (6.5,-3) [thick] {$z_3$};
 \node at (8.5,-3) [thick] {$z_4$};
 \node at (4,-4.5) [thick] {$2$};
 \node at (6,-4.5) [thick] {$3$};
 \node at (8,-4.5) [thick] {$4$};
 \node at (-0.5,0) [thick] {$0$};
 \node at (1,0.5) [thick] {$x$};
 \node at (10.5,-2) [thick] {$1$};
 \node at (9,-1.5) [thick] {$z_1$};
 \end{tikzpicture}
 \end{center}
 \caption{Starting point for the computation of  $\langle 1101 |M(x|\bar z) | 1110 \rangle$.\label{caneva:ex}}
\end{figure}

\begin{figure}[htb]
\begin{center}
 \begin{tikzpicture}[scale=0.65]
 \draw[->,ultra thick] (0,0) -- (1,0) ; 
 \draw[ultra thick] (1,0) -- (10,0) ;
 \draw[ultra thick] (2,0) -- (2,2) ;
 \draw[->,ultra thick] (4,-4) -- (4,-3) ; \draw[ultra thick] (4,-3) -- (4,2) ;
 \draw[->,ultra thick] (6,-4) -- (6,-3) ; \draw[ultra thick] (6,-3) -- (6,-2) ;
 \draw[ultra thick] (8,-2) -- (8,2) ; 
 \draw[ultra thick] (4,-2) -- (6,-2) ;
 \draw[->,ultra thick] (10,-2) -- (9,-2) ; \draw[ultra thick] (9,-2) -- (8,-2) ;
 \draw[ultra thick] (2,0) arc (180:270:2cm);
 \draw[dashed] (6,-2) -- (6,2);
 \draw[->,dashed] (8,-4) -- (8,-3);\draw[dashed] (8,-3) -- (8,-2);
 \draw[dashed] (6,-2) -- (8,-2);
 \draw (9.8,-0.2) --(10.2,0.2);\draw (9.6,-0.2) --(10,0.2);
 \draw (-0.2,-0.2) --(0.2,0.2);\draw (0,-0.2) --(0.4,0.2);
 \node at (4.5,-3) [thick] {$z_2$};
 \node at (6.5,-3) [thick] {$z_3$};
 \node at (8.5,-3) [thick] {$z_4$};
 \node at (4,-4.5) [thick] {$2$};
 \node at (6,-4.5) [thick] {$3$};
 \node at (8,-4.5) [thick] {$4$};
 \node at (-0.5,0) [thick] {$0$};
 \node at (1,0.5) [thick] {$x$};
 \node at (10.5,-2) [thick] {$1$};
 \node at (9,-1.5) [thick] {$z_1$};
 \node at (6,-5.5) [thick] {$(1-\frac{x}{z_3})\frac{z_3}{z_1}$};
 
 \draw[->,ultra thick] (12,0) -- (13,0) ; 
 \draw[ultra thick] (13,0) -- (16,0) ;
 \draw[ultra thick] (14,0) -- (14,2) ;
 \draw[->,ultra thick] (16,-4) -- (16,-3) ; \draw[ultra thick] (16,-3) -- (16,-2) ;
 \draw[->,ultra thick] (18,-4) -- (18,-3) ; \draw[ultra thick] (18,-3) -- (18,0) ;
 \draw[ultra thick] (20,-2) -- (20,2) ; 
 \draw[ultra thick] (18,0) -- (22,0) ;
 \draw[ultra thick] (16,0) -- (16,2) ;
 \draw[->,ultra thick] (22,-2) -- (21,-2) ; \draw[ultra thick] (21,-2) -- (20,-2) ;
 \draw[ultra thick] (14,0) arc (180:270:2cm);
 \draw[dashed] (18,0) -- (18,2);
 \draw[->,dashed] (20,-4) -- (20,-3);\draw[dashed] (20,-3) -- (20,-2);
 \draw[dashed] (16,-2) -- (20,-2);
 \draw[dashed] (16,-2) -- (16,0);
 \draw[dashed] (16,0) -- (18,0);
 \draw (21.8,-0.2) --(22.2,0.2);\draw (21.6,-0.2) --(22,0.2);
 \draw (11.8,-0.2) --(12.2,0.2);\draw (12,-0.2) --(12.4,0.2);
 \node at (16.5,-3) [thick] {$z_2$};
 \node at (18.5,-3) [thick] {$z_3$};
 \node at (20.5,-3) [thick] {$z_4$};
 \node at (16,-4.5) [thick] {$2$};
 \node at (18,-4.5) [thick] {$3$};
 \node at (20,-4.5) [thick] {$4$};
 \node at (11.5,0) [thick] {$0$};
 \node at (13,0.5) [thick] {$x$};
 \node at (22.5,-2) [thick] {$1$};
 \node at (21,-1.5) [thick] {$z_1$};
  \node at (18,-5.5) [thick] {$(1-\frac{z_3}{z_1})\frac{x}{z_1}$};
 \end{tikzpicture}
 \caption{The two different drawings involved in the computation of the transition rate
 $\langle 1101 |M(x|\bar z) | 1110 \rangle$ with their respective weights.
 \label{toto}}
 \end{center}
\end{figure}
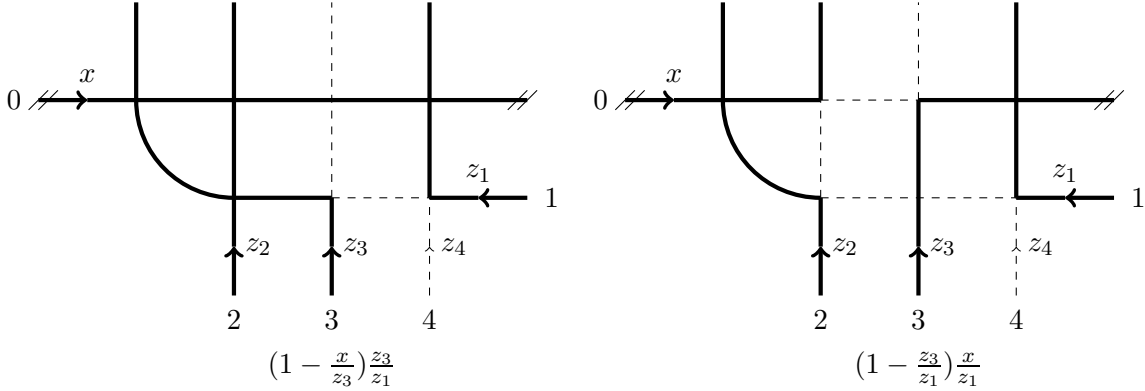

As an example we take $L=4$ and use the  graphical interpretation  to compute the transition rate between the initial configuration $(1,1,1,0)$ and the 
final configuration $(1,1,0,1)$. The initial (resp. final) configuration fixes the form of the incoming (resp. out-going) external lines
(dashed or thick) as in fig. \ref{caneva:ex}.
 Then, we look for drawings of the form given in fig. \ref{caneva:ex}
where the remaining thin lines have to be replaced by thick or dashed lines in such a way that  the weights (as given in fig. \ref{matrix_elements})
of all the vertices do not vanish. The total weight of a given possible drawing is then the product of all these weights.
It is easy to see that there are only two possible drawings, given in fig. \ref{toto} together with their corresponding weights.
Finally the weight of $\langle 1101 |M(x|\bar z) | 1110 \rangle$ is the sum of the weights of the possible drawings.\\

Using this graphical interpretation, we are able to compute all the possible rates between
 any two  configurations. We remark in particular that 
the number of particles is conserved (as mentioned previously each non-vanishing vertex preserves the number of particles). 
Therefore, we restrict ourselves to a given sector with a fixed number of particles. We can also show that all the rates starting from a 
given configuration (with at least one particle) sum to one which proves that $M(x|\bar z)$ can be used as a Markov  matrix in 
the master equation \eqref{eq:MEv}. One has to impose also
\begin{equation}\label{eq:cond}
0\leq \frac{z_i}{z_1}\leq1 \quad\mbox{and}\quad 0\leq \frac{x}{z_i}\leq1 \,,\quad i=1,2,...L,
\end{equation}
so that the probabilities are positive and less than 1. The sector with no particle
is special: its  dimension is $1$ and the matrix $M(x|\bar z)$ is reduced to the scalar 
 $1+\prod_{i=1}^{L}(1-x/z_i)$. Therefore, it cannot be considered as a Markov matrix in the empty sector.
From now, we consider only the cases with at least one particle.

\subsection{Stochastic updating rules \label{subsec:su}} 
 
 The Markov process given by $M(x|\bar z)$ can be  interpreted as a discrete time process with sequential update.
The configuration at the time $t+1$ is obtained from the one at time $t$ by the following dynamics:
\begin{itemize}
 \item \textbf{Particle update:} starting from  right to  left (i.e. from the site $L$ to the site $L-1$ 
and so on), a particle at the site $i$ jumps to the right on the neighboring site
 with a probability $1-z_i/z_1$ provided this site is empty. The particle does not jump
 with the probability $z_i/z_1$. We remind that we are on a periodic lattice, so that
 the  site on the right of the site $L$ is the site $1$. Note that a particle located on site 1 does not move.

 \item \textbf{Hole update:}  once the particle update is done, one performs the hole update. 
 Contrarily to the particle update, we do not necessarily start  and finish at the sites 1 or  $L$. 
 Let $r$ be the site number of a particle (we recall that we restrict ourselves to the case with at least one particle). Starting from the site $r$, we go from  left to 
  right up to the site $r-1$, using periodicity and knowing that the site on the 
  right of the site $L$ is the site 1.   A hole at the site $i$ jumps to 
 the left on the neighboring site with the probability
  $1-x/z_{i-1}$ provided the site is occupied. The hole may stay at site $i$ with probability $x/z_{i-1}$. 
 By convention, we set $z_0=z_L$.
\end{itemize}
As mentioned previously, we would like to emphasize that, due to the inhomogeneities in the transfer matrix, the rates depend on the site 
where the particle or the hole is  situated.
 All the probabilities are positive and less than 1 thanks to \eqref{eq:cond}.
Let us also mention that for the homogeneous case, the update simplifies.
 Indeed, the first step becomes trivial: the particles do not move.

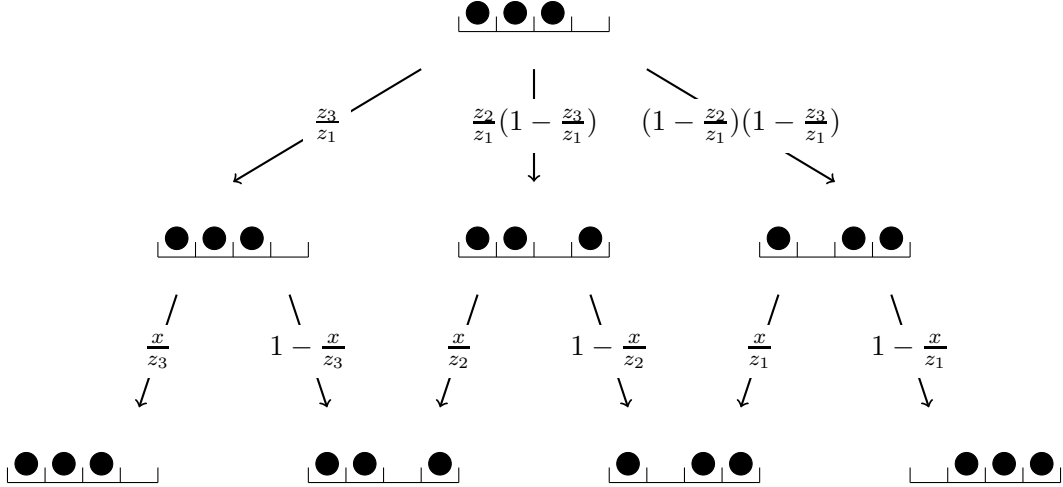
\begin{figure}[htb]
\begin{center}
 \begin{tikzpicture}[scale=1]
  \draw (5,6) -- (7,6) ;
\foreach \i in {5,5.5,...,7}
{\draw (\i,6) -- (\i,6.2) ;}
\draw  (5.25,6.25) circle (0.15) [fill,circle] {};
\draw  (5.75,6.25) circle (0.15) [fill,circle] {};
\draw  (6.25,6.25) circle (0.15) [fill,circle] {};

\draw[->,thick] (4.5,5.5)--  node [fill=white] {$\frac{z_3}{z_1}$} (2,4);
\draw[->,thick] (6,5.5)--  node [fill=white] {$\frac{z_2}{z_1}(1-\frac{z_3}{z_1})$}(6,4);
\draw[->,thick] (7.5,5.5)--  node [fill=white] {$(1-\frac{z_2}{z_1})(1-\frac{z_3}{z_1})$}(10,4);

 \draw (1,3) -- (3,3) ;
\foreach \i in {1,1.5,...,3}
{\draw (\i,3) -- (\i,3.2) ;}
\draw  (1.25,3.25) circle (0.15) [fill,circle] {};
\draw  (1.75,3.25) circle (0.15) [fill,circle] {};
\draw  (2.25,3.25) circle (0.15) [fill,circle] {};
 \draw (5,3) -- (7,3) ;
\foreach \i in {5,5.5,...,7}
{\draw (\i,3) -- (\i,3.2) ;}
\draw  (5.25,3.25) circle (0.15) [fill,circle] {};
\draw  (5.75,3.25) circle (0.15) [fill,circle] {};
\draw  (6.75,3.25) circle (0.15) [fill,circle] {};
 \draw (9,3) -- (11,3) ;
\foreach \i in {9,9.5,...,11}
{\draw (\i,3) -- (\i,3.2) ;}
\draw  (9.25,3.25) circle (0.15) [fill,circle] {};
\draw  (10.25,3.25) circle (0.15) [fill,circle] {};
\draw  (10.75,3.25) circle (0.15) [fill,circle] {};

\draw[->,thick] (1.25,2.5)--  node [fill=white] {$\frac{x}{z_3}$} (0.75,1);
\draw[->,thick] (2.75,2.5)--  node [fill=white] {$1-\frac{x}{z_3}$} (3.25,1);
\draw[->,thick] (5.25,2.5)--  node [fill=white] {$\frac{x}{z_2}$} (4.75,1);
\draw[->,thick] (6.75,2.5)--  node [fill=white] {$1-\frac{x}{z_2}$} (7.25,1);
\draw[->,thick] (9.25,2.5)--  node [fill=white] {$\frac{x}{z_1}$} (8.75,1);
\draw[->,thick] (10.75,2.5)--  node [fill=white] {$1-\frac{x}{z_1}$} (11.25,1);

\draw (-1,0) -- (1,0) ;
\foreach \i in {-1,-0.5,...,1}
{\draw (\i,0) -- (\i,0.2) ;}
\draw  (-0.75,0.25) circle (0.15) [fill,circle] {};
\draw  (-0.25,0.25) circle (0.15) [fill,circle] {};
\draw  (0.25,0.25) circle (0.15) [fill,circle] {};
\draw (3,0) -- (5,0) ;
\foreach \i in {3,3.5,...,5}
{\draw (\i,0) -- (\i,0.2) ;}
\draw  (3.25,0.25) circle (0.15) [fill,circle] {};
\draw  (3.75,0.25) circle (0.15) [fill,circle] {};
\draw  (4.75,0.25) circle (0.15) [fill,circle] {};
\draw (7,0) -- (9,0) ;
\foreach \i in {7,7.5,...,9}
{\draw (\i,0) -- (\i,0.2) ;}
\draw  (7.25,0.25) circle (0.15) [fill,circle] {};
\draw  (8.25,0.25) circle (0.15) [fill,circle] {};
\draw  (8.75,0.25) circle (0.15) [fill,circle] {};
\draw (11,0) -- (13,0) ;
\foreach \i in {11,11.5,...,13}
{\draw (\i,0) -- (\i,0.2) ;}
\draw  (11.75,0.25) circle (0.15) [fill,circle] {};
\draw  (12.25,0.25) circle (0.15) [fill,circle] {};
\draw  (12.75,0.25) circle (0.15) [fill,circle] {};
 \end{tikzpicture}
 \caption{An example of sequential update corresponding to the Markov matrix $M(x|\bar z)$. The first line is
 the configuration at time $t$ and the third line shows the possible configurations at  time $t+1$.
 The second line corresponds to the intermediate configurations after the update of the particles and the hole update still to be done.
 The label of the arrows provides the rate of the corresponding change of configurations.
 \label{fig:seq}}
 \end{center}
\end{figure}

 These rules are illustrated in  figure \ref{fig:seq} for a  chain with 4 sites in the configuration $(1,1,1,0)$
at time $t$. 
We deduce from this figure the different possible rates between the configurations
 which correspond  to the entries of $M(x|\bar z)$.
One gets
\begin{equation}
 M(x|\bar z)\big( (1,1,0,1),(1,1,1,0)\big)=\langle 1101 |M(x|\bar z) | 1110 \rangle
=\left(1-\frac{x}{z_3}\right)\frac{z_3}{z_1}+\left(1-\frac{z_3}{z_1}\right)\frac{x}{z_1}
\end{equation}
in accordance with the calculation done in section \ref{subsec:graph} (see also figure \ref{toto}).

\paragraph{Justification of the sequential up-date.} The sequential update described above can be easily identified when considering the Markov matrix $M(x|\bar z)$ at the special point $x=z_j$. Indeed, we write 
$M(z_j|\bar z) = \Big( t(z_j|\bar z)\fC\Big)\Big(\fC^{-1} t(z_1|\bar z)^{-1}\Big)$ where $\fC$ is the cyclic permutation. 
Then, from the explicit expressions
\begin{eqnarray}
t(z_j|\bar z) &=& R_{j,j-1}(\frac{z_j}{z_{j-1}})\cdots R_{j,1}(\frac{z_j}{z_{1}})\,R_{j,L}(\frac{z_j}{z_{L}})\cdots R_{j,j+1}(\frac{z_j}{z_{j+1}})
\\
t(z_1|\bar z)^{-1} &=& R_{2,1}(\frac{z_2}{z_{1}})\, R_{3,1}(\frac{z_{3}}{z_{1}})\cdots R_{L,1}(\frac{z_L}{z_{1}})
\end{eqnarray}
it is easy to see that $\fC^{-1} t(z_1|\bar z)^{-1}$ corresponds to the particle update, while $t(z_j|\bar z)\fC$ corresponds to the hole update 
at $x=z_j$.  

Let us remark that such a simple sequential update is specific to the totally asymmetric exclusion process.
For the partially asymmetric case, it would be much more involved.

\subsection{Stationary State Properties}
\label{subsec:stat}

 We now investigate some properties of the stationary state of the Markov chain  defined by $M(x|\bar z)$.
We  construct the steady state in a formal manner, then give  an elementary expression  for the
stationary  measure. Finally, we obtain   some simple formulas
 for steady-state correlations.

\subsubsection{Construction of the steady state}

The building block of the stationary state is the following vector
\begin{equation} \label{v}
 v(z)=\left(
 \begin{array}{c}
 z\\1
 \end{array}
\right).
\end{equation}
It satisfies the Zamolodchikov-Faddeev relation \cite{ZF}
\begin{equation} \label{eq:ZF_scalar}
 R_{12}(z_1/z_2)v_1(z_1)v_2(z_2)=v_1(z_1)v_2(z_2) \, . 
\end{equation}
The stationary state of the process $M(x|\bar z)$ is constructed from 
\begin{equation} \label{steady_state}
 \steady = v_1(z_1)v_2(z_2)\dots v_L(z_L).
\end{equation}
Recall that the subscripts denote which component of the tensor space the vector $v$ belongs to.
To be more precise, since the process conserves the number of particles, there are $L$ independent sectors
(as mentioned previously we do not consider the empty sector), each corresponding to a given number $m$
of particles  in the system. The stationary state is hence degenerate: there is one stationary state for each sector. 
Moreover, the exact normalization $Z^{(m)}$ of the stationary state depends on the sector we are considering.  Then, each stationary state is given
by the components of $\steady$  corresponding to the sector and correctly normalized. 

The following calculation justifies that $\steady$ given in \eqref{steady_state}
 is the stationary state  of the system. Indeed, for $i=1,2,...,L$ we have
\begin{eqnarray*}
 t(z_i|\bar z) \steady & = & R_{i,i-1}(\frac{z_i}{z_{i-1}})\dots R_{i,1}(\frac{z_i}{z_{1}})R_{i,L}(\frac{z_i}{z_L})\dots R_{i,i+1}(\frac{z_i}{z_{i+1}})v_1(z_1)\dots v_L(z_L)\\
 & = & v_1(z_1)v_2(z_2)\dots v_L(z_L) \\
 & = & \steady .
\end{eqnarray*}
The first equality is obtained using the regularity property of the R-matrix \eqref{R-regularity} whereas the second equality is obtained using $L-1$
times the property \eqref{eq:ZF_scalar}. 
Clearly, $t(x|\bar z)$ is a polynomial of degree less or equal to $L$ in $x$. It is possible to show (using  the graphical interpretation  for instance)
that in the sectors where there is at least one particle, {the degree of $t(x|\bar z)$ is in fact at most equal to $L-1$.} Hence we can deduce through 
interpolation arguments that in these sectors, 
the stationary state is given by the corresponding components of $\steady$. 
Note  that when we take the homogeneous limit $z_i \rightarrow 1$ we recover
 the  uniform stationary distribution of the continuous time TASEP.

\subsubsection{Calculation of steady-state averages}

From the knowledge of the stationary distribution, we can calculate various  physical quantities. 
We shall see that some observables can be expressed as symmetric polynomials in the inhomogeneity
parameters $z_1,\ldots,z_L$.
  
\paragraph{Probability weight of a given configuration.}
The weight of the configuration $(\tau_1,\dots,\tau_L)$ is readily obtained  using \eqref{steady_state}: 
\begin{equation}
 \mathcal{S}(\tau_1,\dots,\tau_L)=\prod_{i=1}^{L}\left(\delta_{1,\tau_i}+z_i\delta_{0,\tau_i}\right).
\end{equation}

\paragraph{Normalization factor.}
In the sector with $m$ particles, the normalization factor of the stationary state is obtained by summing the weights of all the configurations
with $m$ particles
\begin{equation}\label{eq:ZZ}
 Z^{(m)}=\sum\limits_{\substack{ I \subset  \{1,\dots,L\} \\ |I|=L-m}}^{}\ {\prod_{i \in I} z_i \ =\ e_{L-m}(z_1,\dots,z_L)} \, ,
\end{equation}
 where $e_{L-m}$ is the elementary symmetric homogeneous polynomial of degree $L-m$. The normalization factor can be written as a  Schur polynomial:
\begin{equation}
 Z^{(m)}=s^A_{1^{L-m}}(z_1,\dots,z_L),\quad \mbox{where}\quad 1^{L-m}=(\underbrace{1,...,1}_{L-m},\underbrace{0,...,0}_{m}).
\end{equation}
We remind the definition of the Schur polynomial (of type A) associated with 
 a partition  $\lambda=(\lambda_1,\dots,\lambda_L)$ with $\lambda_1\geq\dots\geq\lambda_L\geq0$:
\begin{equation}
s^A_{\lambda}(z_1,\dots,z_L)=
\frac{\det \left((z_j)^{L-i+\lambda_i} \right)_{i,j}}{\det \left(z_j^{L-i} \right)_{i,j}}.
\end{equation}
This expression of   the partition function in terms of the Schur polynomial allows us to relate the value of $Z^{(m)}$ in the homogeneous limit ($z_i\rightarrow 1$) 
with the dimension of the representation $\pi^A(\lambda)$ of $sl(L)$ labeled by the Young tableau $[\lambda]$.
 Indeed, using the Weyl character formula (see e.g. \cite{GTM,dico} for a review), we obtain 
\begin{equation}\label{eq:rsl1}
s^A_{\lambda}(1,\dots,1)=\prod_{1\leq j\leq i\leq L} \frac{\lambda_j-\lambda_i+i-j}{i-j}=\text{dim}(\pi^A(\lambda))\;.
\end{equation}
In particular, we have 
\begin{equation}\label{eq:rsl2}
 Z^{(m)}\Big|_{z_1=\dots=z_L=1}=\text{dim}(\pi^A(1^{L-m}))=\left(\begin{array}{c} L \\m\end{array}\right)
\end{equation}
in accordance with a direct computation starting from \eqref{eq:ZZ}. 

\paragraph{Density.}
In the sector with $m$ particles, the particle density at site $i$ is obtained by summing the weights of all the configurations with $m$ particles,
one of them being at site $i$
\begin{equation} \label{density_periodic}
 \langle n_i \rangle =\frac{e_{L-m}(z_1,\dots,z_{i-1},z_{i+1},\dots,z_L)}{e_{L-m}(z_1,\dots,z_L)}\;.
\end{equation}
We can show that $\displaystyle \sum_{i=1}^L \langle n_i \rangle=m$ as expected.

\paragraph{Higher correlation functions.}

  The higher correlation functions take also a very simple form. The correlations between the sites $i_1<i_2<\dots<i_\ell$ is given by
\begin{equation} \label{corr_periodic}
 \langle n_{i_1}n_{i_2}\dots n_{i_\ell} \rangle =\frac{e_{L-m}(z_1,\dots,z_{i_1-1},z_{i_1+1},\dots,z_{i_2-1},z_{i_2+1},\dots,z_{i_\ell-1},z_{i_\ell+1},\dots,z_L)}{e_{L-m}(z_1,\dots,z_L)}\;.
\end{equation}
For $\ell>m$, the correlation functions vanish as expected since the number of particles is $m$ and the correlation
functions for more than $m$ particles has no meaning.

%
%
%

\section{Markov  process on the open lattice with boundaries \label{sec:op}}

 We now consider a model  on a finite lattice with  two boundaries
 that play the role of 
reservoirs from which particles can enter the system  or exit from it.
 In order to construct an integrable system, one needs  to solve simultaneously
 the Yang-Baxter equation in the bulk, leading to  the $R$-matrix,  and
 Sklyanin's reflection equations at the boundaries \cite{sklyanin}, which
 define two boundary operators $K$ and $\widetilde{K}$.
 These operators  allow us to
 construct an inhomogeneous transfer matrix  in section \ref{subsec: InhOPEN}, that 
 we illustrate  graphically in section
  \ref{subsec: GraphicOPEN} and  interpret as 
 a stochastic process in section \ref{subsec: rulesOPEN}. 
 The  stationary state of this process is computed 
 using the matrix product technique  \cite{DEHP,MartinRev} that 
 allows us to calculate stationary observables (section
\ref{subsec:StatOpen}). Using the explicit expression of the normalisation function, we relate
Bethe roots and Lee-Yang zeros in section \ref{sec:bly}.

\subsection{Inhomogeneous transfer matrix for the open system}
 \label{subsec: InhOPEN}

The building blocks of the transfer matrices for the open
 system  are the $R$ matrix defined in section \ref{R-TASEP} and
the following boundary  matrices: 
\begin{equation}\label{eq:K}
 K(x)=\left(
 \begin{array}{cc}
 \frac{(a+x)x}{xa+1} & 0\\ \frac{1-x^2}{xa+1} & 1
 \end{array}
\right) \,\,\,\, \hbox{  and } \,\,\,\, 
\widetilde{K}(x)=\left(
 \begin{array}{cc}
 \frac{1}{xb+1} &\frac{1}{xb+1}\\ 0 & \frac{xb}{xb+1}
 \end{array}
\right).
\end{equation}
These matrices act in the space $\mathbb{C}^2$ and we choose the basis $|0\rangle$, $|1\rangle$ corresponding to the lattice configurations
$(0)$, $(1)$ respectively. These $K$-matrices are associated with the continuous time TASEP with boundaries as will be explained below. 

The matrix $K$ obeys the reflection equation  \cite{sklyanin}
\begin{equation}
\label{eq:re}
 R_{12}(\frac{x_1}{x_2})\, K_1(x_1)\, R_{21}(x_1 x_2)\, K_2(x_2)=
 K_2(x_2)\,R_{12}(x_1 x_2)\,K_1(x_1)\,R_{21}(\frac{x_1}{x_2})\;.
\end{equation}
The matrix $\widetilde K$ is related to an auxiliary $\bar K$ matrix,
 defined as 
$\bar K(x)= 
 tr_0 \left(\widetilde{K}_0(\frac{1}{x})R_{01}(\frac{1}{x^2})P_{01}\right),$ 
which obeys the  reflection equation \eqref{eq:re} associated
 with the matrix $\bar R_{12}(x)=R_{21}(1/x)$:
 \begin{equation} \label{eq:barre}
\bar  R_{12}(\frac{x_1}{x_2})\, \bar{K_1}(x_1)\, \bar R_{21}({x_1 x_2})\, \bar{K_2}(x_2)=
 \bar{K_2}(x_2)\,\bar R_{12}({x_1 x_2}) \,  \bar{K_1}(x_1)\,\bar R_{21}(\frac{x_1}{x_2})\;.
\end{equation}
 The explicit form of   $\bar K$  is 
\begin{equation}
\bar K(x)= \left( \begin {array}{cc} 
1&\displaystyle{\frac{(x^2-1)}{x(x+b)}}\\
0&\displaystyle{\frac{xb+1}{x(x+b)}}
\end {array} \right).
\end{equation}

  We can now construct the  inhomogeneous open transfer matrix
 as follows: 
\begin{equation} \label{eq:transfer_matrix}
t(x|\bar z) = tr_0\Big( \widetilde K_0(x)\,R_{0,L}(\frac{x}{z_L})\dots R_{0,1}(\frac{x}{z_1})\,K_0(x)\,R_{1,0}(xz_1)\dots R_{L,0}(xz_L) \Big).
\end{equation}
It is shown in \cite{CRV} that these  transfer matrices  commute
 for different values of the spectral parameter: $[t(x|\bar z),t(y|\bar z)]=0$.
Note that the derivation of this property cannot be obtained directly from the formalism developed in \cite{sklyanin}, because the matrix $R$ does not 
satisfy  the crossing unitarity.
{We use the operator
 $t(x|\bar z)$  to define the following discrete time Markov  process
 \begin{equation}
  |P_{t+1}\rangle=t(x|\bar z)|P_{t}\rangle\;.
 \end{equation}
 The parameters must satisfy the following constraints
 \begin{equation}
  0\leq xz_i\leq 1,\quad0\leq \frac{x}{z_i}\leq 1\quad\text{and}\qquad ax,bx\geq 0
 \end{equation}
to ensure that the entries of $t(x|\bar z)$ are probabilities. We can also show that the entries on each column of $t(x|\bar z)$ sum to one, which guarantees
the conservation of the probability $|P_{t}\rangle$.}

\paragraph{Continuous time limit.}
 In  the homogeneous limit $z_i \rightarrow 1$, we are left with the operator $t(x)=\left. t(x|\bar z) \right|_{z_i=1}$. 
A  straightforward computation gives  
\begin{equation}
 -\frac{1}{2} t'(1)= -\frac{1}{2}K_1'(1)-\sum_{k=1}^{L-1}P_{k,k+1}.R_{k,k+1}'(1)+\frac{1}{2}\bar K_L'(1)\;.
\end{equation}
where the prime $(\,.\,)'$ stands for the derivative with respect to the spectral parameter.
 Then, introducing the parameters $\alpha=1/(a+1)$ and $\beta=1/(b+1)$,
we obtain 
\begin{equation}
-\frac{1}{2}K'(1)=\left(\begin{array}{cc}
                         -\alpha & 0 \\
                         \alpha & 0
                        \end{array}\right), \quad 
 -P.R'(1)=\left(\begin{array}{cccc}
           0&0&0&0\\
           0&0&1&0\\
           0&0&-1&0\\
           0&0&0&0
          \end{array}\right), \quad 
 \frac{1}{2}\bar K'(1)=\left(\begin{array}{cc}
                         0 & \beta \\
                         0 & -\beta
                        \end{array}\right)\;.
\end{equation}
Hence $-1/2\,t'(1)$ is  nothing but
 the continuous time Markov  matrix of the open TASEP.
 Therefore, the stationary state of the continuous time TASEP computed in \cite{DEHP}
can be   obtained by taking   the
 homogeneous limit of the one computed here.

\subsection{Graphical representation  of the transition rates}
\label{subsec: GraphicOPEN}

\begin{figure}[htb]
\begin{center}
 \begin{tikzpicture}[scale=0.7]
\foreach \i in {-6,0,6,12}
{\draw (\i,-1.5) -- (\i,1.5) ;
\foreach \j in {-1.2,-0.9,...,1.6}
{\draw (\i-0.3,\j-0.3) -- (\i,\j);} ;}
\foreach \i in {-6,0}
{\draw[->,dashed] (\i+1.5,-1.5) -- (\i+0.75,-0.75) ; \draw[dashed] (\i+0.75,-0.75) -- (\i,0) ;}
\foreach \i in {6,12}
{\draw[->,ultra thick] (\i+1.5,-1.5) -- (\i+0.75,-0.75) ; \draw[ultra thick] (\i+0.75,-0.75) -- (\i,0) ;}
\foreach \i in {-6,6}
{\draw[->,dashed] (\i,0) -- (\i+0.75,0.75) ; \draw[dashed] (\i+0.75,0.75) -- (\i+1.5,1.5) ;}
\foreach \i in {0,12}
{\draw[->,ultra thick] (\i,0) -- (\i+0.75,0.75) ; \draw[ultra thick] (\i+0.75,0.75) -- (\i+1.5,1.5) ;}
\foreach \i in {-6,0,6,12}
{\node at (\i+1.4,-0.75) [] {\footnotesize{$1/x$}};   }
\foreach \i in {-6,0,6,12}
{\node at (\i+0.3,0.75) [] {\footnotesize{$x$}};   }
\node at (-5.5,-2.5) [] {$\frac{(a+x)x}{xa+1}$} ;
\node at (0.5,-2.5) [] {$\frac{1-x^2}{xa+1}$} ;
\node at (6.5,-2.5) [] {$0$} ;
\node at (12.5,-2.5) [] {$1$} ;
\node at (-5.5,2.5) [] {$\langle0| K(x)|0\rangle$} ;
\node at (0.5,2.5) [] {$\langle1| K(x)|0\rangle$} ;
\node at (6.5,2.5) [] {$\langle0| K(x)|1\rangle$} ;
\node at (12.5,2.5) [] {$\langle1| K(x)|1\rangle$} ;
 \end{tikzpicture}
 \caption{Graphical representation of the $K$-matrix. \label{fig:Kmat}}
 \end{center}
\end{figure}
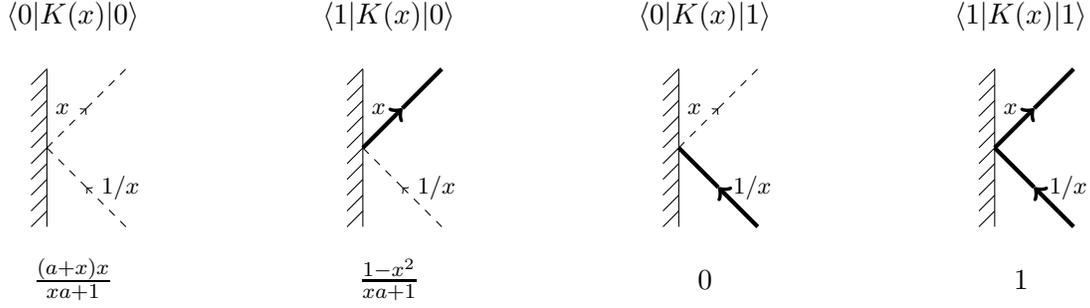

In fig. \ref{fig:Kmat}, each  matrix element of the $K$-matrix is drawn
as a left-reflection diagram.  
The incoming lines (according to the arrows direction)
 before the reflection point denote the vector the matrix $K$ is acting on.
A dashed line corresponds to
 the vector $|0\rangle$ (or equivalently  to an empty site),
whereas a continuous thick line   corresponds to   $|1\rangle$ 
 (or equivalently  to an occupied site).
In a similar way, the out-going lines (after the reflection point)
 represent  the state of the vector with which we are contracting to the
left the matrix $K$.

\begin{figure}[htb]
\begin{center}
 \begin{tikzpicture}[scale=0.7]
\foreach \i in {-6,0,6,12}
{\draw (\i,-1.5) -- (\i,1.5) ;
\foreach \j in {-1.2,-0.9,...,1.6}
{\draw (\i+0.3,\j-0.3) -- (\i,\j);} ;}
\foreach \i in {-6,6}
{\draw[->,dashed] (\i,0) -- (\i-0.75,-0.75) ; \draw[dashed] (\i-0.75,-0.75) -- (\i-1.5,-1.5) ;}
\foreach \i in {0,12}
{\draw[->,ultra thick] (\i,0) -- (\i-0.75,-0.75) ; \draw[ultra thick] (\i-0.75,-0.75) -- (\i-1.5,-1.5) ;}
\foreach \i in {-6,0}
{\draw[->,dashed] (\i-1.5,1.5) -- (\i-0.75,0.75) ; \draw[dashed] (\i-0.75,0.75) -- (\i,0) ;}
\foreach \i in {6,12}
{\draw[->,ultra thick] (\i-1.5,1.5) -- (\i-0.75,0.75) ; \draw[ultra thick] (\i-0.75,0.75) -- (\i,0) ;}
\foreach \i in {-6,0,6,12}
{\node at (\i-1.4,-0.75) [] {\footnotesize{$1/x$}};   }
\foreach \i in {-6,0,6,12}
{\node at (\i-0.3,0.75) [] {\footnotesize{$x$}};   }
\node at (-6.5,-2.5) [] {$\frac{1}{xb+1}$} ;
\node at (-0.5,-2.5) [] {$0$} ;
\node at (5.5,-2.5) [] {$\frac{1}{xb+1}$} ;
\node at (11.5,-2.5) [] {$\frac{xb}{xb+1}$} ;
\node at (-6.5,2.5) [] {$\langle0|\widetilde K(x)|0\rangle$} ;
\node at (-0.5,2.5) [] {$\langle0|\widetilde K(x)|1\rangle$} ;
\node at (5.5,2.5) [] {$\langle1|\widetilde K(x)|0\rangle$} ;
\node at (11.5,2.5) [] {$\langle1|\widetilde K(x)|1\rangle$} ;
 \end{tikzpicture}
 \caption{Graphical representation of the $\widetilde K$-matrix. \label{fig:Ktilde}}
 \end{center}
\end{figure}
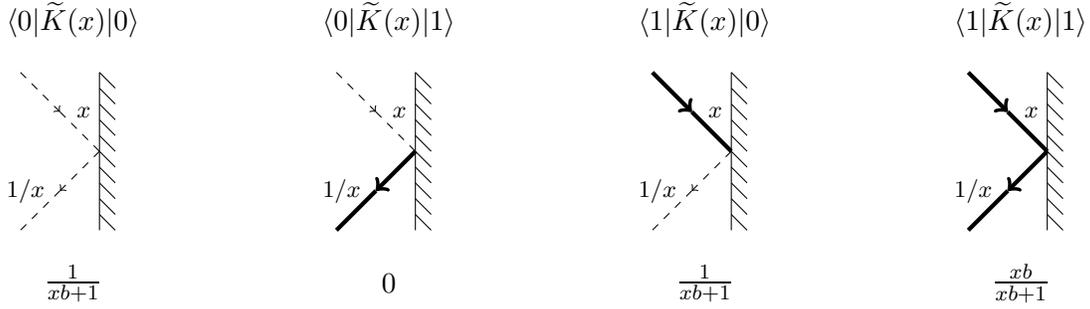

 Similarly, the  graphical representation for the matrix $\widetilde{K}$
 is given  in fig. \ref{fig:Ktilde}.

\begin{figure}[htb]
\begin{center}
 \begin{tikzpicture}[scale=0.7]
\draw (-6,-2) -- (-6,2) ;
\draw (12,-2) -- (12,2) ;
\foreach \i in {-1.6,-1.2,...,2.1}
{\draw (-6-0.4,\i-0.4) -- (-6,\i) ; \draw (12+0.4,\i-0.4) -- (12,\i) ;}
\draw[->] (-4,-2) -- (-5,-1) ; \draw[] (-5,-1) -- (-6,0) ; 
\draw[->] (-6,0) -- (-5,1) ; \draw[] (-5,1) -- (-4,2) ; 
\draw[->] (10,2) -- (11,1) ; \draw[] (11,1) -- (12,0) ; 
\draw[->] (12,0) -- (11,-1) ; \draw[] (11,-1) -- (10,-2) ;
\draw[->] (10,-2) -- (9.5,-2) ; \draw[->] (9.5,-2) -- (-3.5,-2) ; \draw[] (-3.5,-2) -- (-4,-2) ;
\draw[->] (-4,2) -- (-3.5,2) ; \draw[->] (-3.5,2) -- (9.5,2) ; \draw[] (9.5,2) -- (10,2) ;
\foreach \i in {-3,-1,...,10}
{\draw[->] (\i,-3) -- (\i,-2.5) ; \draw[->] (\i,-2.5) -- (\i,1.5) ; \draw[] (\i,1.5) -- (\i,3) ; }
\node at (-5.5,1) [] {$x$} ; \node at (11.5,1) [] {$x$} ;
\node at (-4.25,-1) [] {$1/x$} ; \node at (10.25,-1) [] {$1/x$} ;
\node at (-2.5,-2.5) [] {$z_1$} ; \node at (-0.5,-2.5) [] {$z_2$} ; \node at (3.5,-2.5) [] {$z_i$} ;
\node at (7.75,-2.5) [] {$z_{L-1}$} ; \node at (9.5,-2.5) [] {$z_L$} ;
\node at (-3,-3.5) [] {$1$} ; \node at (-1,-3.5) [] {$2$} ; \node at (3,-3.5) [] {$i$} ;
\node at (7,-3.5) [] {$L-1$} ; \node at (9,-3.5) [] {$L$} ;
\node at (1,-3.5) [] {$\dots$} ; \node at (5,-3.5) [] {$\dots$} ;
\node at (-4.5,2) [] {$0$} ; \node at (10.5,-2) [] {$0$} ;
 \end{tikzpicture}
 \caption{Graphical representation of the transfer matrix.\label{fig:open-trans}}
 \end{center}
\end{figure}
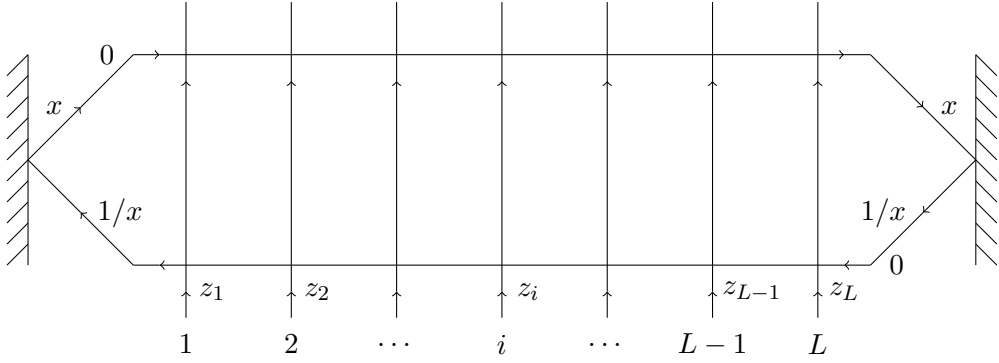

We can now draw diagrams to represent the action of the full transfer matrix.
 The general picture is  displayed in fig. \ref{fig:open-trans}. 

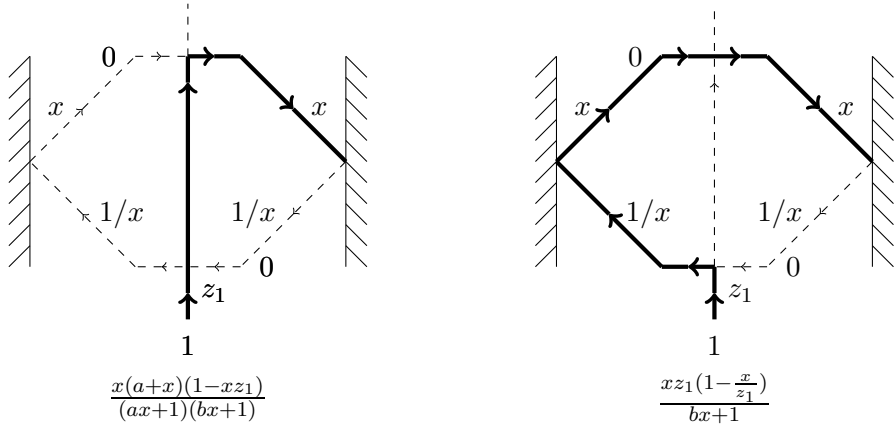
\begin{figure}[htb]
\begin{center}
 \begin{tikzpicture}[scale=0.7]
\draw (0,-2) -- (0,2) ;
\draw (6,-2) -- (6,2) ;
\foreach \i in {-1.6,-1.2,...,2.1}
{\draw (-0.4,\i-0.4) -- (0,\i) ; \draw (6+0.4,\i-0.4) -- (6,\i) ;}
\draw[->,dashed] (2,-2) -- (1,-1) ; \draw[dashed] (1,-1) -- (0,0) ; 
\draw[->,dashed] (0,0) -- (1,1) ; \draw[dashed] (1,1) -- (2,2) ; 
\draw[->,ultra thick] (4,2) -- (5,1) ; \draw[ultra thick] (5,1) -- (6,0) ; 
\draw[->,dashed] (6,0) -- (5,-1) ; \draw[dashed] (5,-1) -- (4,-2) ;
\draw[->,dashed] (4,-2) -- (3.5,-2) ; \draw[->,dashed] (3.5,-2) -- (2.5,-2) ; \draw[dashed] (2.5,-2) -- (2,-2) ;
\draw[->,dashed] (2,2) -- (2.5,2) ; \draw[dashed] (2.5,2) -- (3,2) ;\draw[->,ultra thick] (3,2) -- (3.5,2) ; \draw[ultra thick] (3.5,2) -- (4,2) ;
\foreach \i in {3}
{\draw[->,ultra thick] (\i,-3) -- (\i,-2.5) ; \draw[->,ultra thick] (\i,-2.5) -- (\i,1.5) ; \draw[ultra thick] (\i,1.5) -- (\i,2) ; \draw[dashed] (\i,2) -- (\i,3) ;}
\node at (0.5,1) [] {$x$} ; \node at (5.5,1) [] {$x$} ;
\node at (1.75,-1) [] {$1/x$} ; \node at (4.25,-1) [] {$1/x$} ;
\node at (3.5,-2.5) [] {$z_1$} ; 
\node at (3,-3.5) [] {$1$} ; 
\node at (1.5,2) [] {$0$} ; \node at (4.5,-2) [] {$0$} ;
\node at (3.5,-2.5) [] {$z_1$} ; 
\node at (3,-3.5) [] {$1$} ; 
\node at (1.5,2) [] {$0$} ; \node at (4.5,-2) [] {$0$} ;
\node at (3,-4.5) [] {$\frac{x(a+x)(1-xz_1)}{(ax+1)(bx+1)}$} ;

\draw (10,-2) -- (10,2) ;
\draw (16,-2) -- (16,2) ;
\foreach \i in {-1.6,-1.2,...,2.1}
{\draw (10-0.4,\i-0.4) -- (10,\i) ; \draw (16+0.4,\i-0.4) -- (16,\i) ;}
\draw[->,ultra thick] (12,-2) -- (11,-1) ; \draw[ultra thick] (11,-1) -- (10,0) ; 
\draw[->,ultra thick] (10,0) -- (11,1) ; \draw[ultra thick] (11,1) -- (12,2) ; 
\draw[->,ultra thick] (14,2) -- (15,1) ; \draw[ultra thick] (15,1) -- (16,0) ; 
\draw[->,dashed] (16,0) -- (15,-1) ; \draw[dashed] (15,-1) -- (14,-2) ;
\draw[->,dashed] (14,-2) -- (13.5,-2) ;\draw[dashed] (13.5,-2) -- (13,-2) ; \draw[->,ultra thick] (13,-2) -- (12.5,-2) ; \draw[ultra thick] (12.5,-2) -- (12,-2) ;
\draw[->,ultra thick] (12,2) -- (12.5,2) ; \draw[->,ultra thick] (12.5,2) -- (13.5,2) ; \draw[ultra thick] (13.5,2) -- (14,2) ;
\foreach \i in {13}
{\draw[->,ultra thick] (\i,-3) -- (\i,-2.5) ; \draw[ultra thick] (\i,-2.5) -- (\i,-2) ;\draw[->,dashed] (\i,-2) -- (\i,1.5) ; \draw[dashed] (\i,1.5) -- (\i,3) ; }
\node at (10.5,1) [] {$x$} ; \node at (15.5,1) [] {$x$} ;
\node at (11.75,-1) [] {$1/x$} ; \node at (14.25,-1) [] {$1/x$} ;
\node at (13.5,-2.5) [] {$z_1$} ; 
\node at (13,-3.5) [] {$1$} ; 
\node at (11.5,2) [] {$0$} ; \node at (14.5,-2) [] {$0$} ;
\node at (13,-4.5) [] {$\frac{xz_1(1-\frac{x}{z_1})}{bx+1}$} ;
 \end{tikzpicture}
 \caption{Graphical computation of the transition rate
 $\langle 0|t(x|\bar z)|1\rangle$ for $L=1$.
 The two different contributions are represented with
 their respective weights.\label{fig:open-L=1}}
 \end{center}
\end{figure}

As an example we can compute graphically for $L=1$ the transition rate $\langle 0|t(x|\bar z)|1\rangle$
between the initial configuration $(1)$ and the final configuration $(0)$
 (see fig.~\ref{fig:open-L=1}). The sum of both contributions drawn  in fig.~\ref{fig:open-L=1} gives
\begin{equation} \label{weight_example_open}
\langle 0|t(x|\bar z)|1\rangle  =  \frac{x(a+x)(1-xz_1)}{(ax+1)(bx+1)}+\frac{xz_1(1-\frac{x}{z_1})}{bx+1} 
 =  (1-x^2)\frac{x(a+z_1)}{(ax+1)(bx+1)}.
\end{equation}


\clearpage

\subsection{Dynamical rules} 
\label{subsec: rulesOPEN}

\begin{figure}[htbp]
\begin{center}
 \begin{tikzpicture}[scale=0.85]
\draw (9.8,9.8) -- (10.2,9.8) ;\draw (9.8,9.8) -- (9.8,10) ;\draw (10.2,9.8) -- (10.2,10) ;
\draw  (10,10) circle (0.125) [fill,circle] {};

\draw (2.3,6.3) -- (2.7,6.3) ;\draw (2.3,6.3) -- (2.3,6.5) ;\draw (2.7,6.3) -- (2.7,6.5) ;
\draw[dashed] (1.9,6.3) -- (2.3,6.3) ;\draw[dashed] (1.9,6.3) -- (1.9,6.5) ;
\draw[dashed] (2.7,6.3) -- (3.1,6.3) ;\draw[dashed] (3.1,6.3) -- (3.1,6.5) ;
\draw  (2.5,6.5) circle (0.125) [fill,circle] {};

\draw (6.9,6.3) -- (7.3,6.3) ;\draw (6.9,6.3) -- (6.9,6.5) ;\draw (7.3,6.3) -- (7.3,6.5) ;
\draw[dashed] (6.5,6.3) -- (6.9,6.3) ;\draw[dashed] (6.5,6.3) -- (6.5,6.5) ;
\draw[dashed] (7.3,6.3) -- (7.7,6.3) ;\draw[dashed] (7.7,6.3) -- (7.7,6.5) ;
\draw  (7.1,6.5) circle (0.125) [fill,circle] {};
\draw  (7.5,6.5) circle (0.125) [fill,circle] {};

\draw (12.7,6.3) -- (13.1,6.3) ;\draw (12.7,6.3) -- (12.7,6.5) ;\draw (13.1,6.3) -- (13.1,6.5) ;
\draw[dashed] (12.3,6.3) -- (12.7,6.3) ;\draw[dashed] (12.3,6.3) -- (12.3,6.5) ;
\draw[dashed] (13.1,6.3) -- (13.5,6.3) ;\draw[dashed] (13.5,6.3) -- (13.5,6.5) ;
\draw  (12.5,6.5) circle (0.125) [fill,circle] {};
\draw  (12.9,6.5) circle (0.125) [fill,circle] {};

\draw (17.3,6.3) -- (17.7,6.3) ;\draw (17.3,6.3) -- (17.3,6.5) ;\draw (17.7,6.3) -- (17.7,6.5) ;
\draw[dashed] (16.9,6.3) -- (17.3,6.3) ;\draw[dashed] (16.9,6.3) -- (16.9,6.5) ;
\draw[dashed] (17.7,6.3) -- (18.1,6.3) ;\draw[dashed] (18.1,6.3) -- (18.1,6.5) ;
\draw  (17.1,6.5) circle (0.125) [fill,circle] {};
\draw  (17.5,6.5) circle (0.125) [fill,circle] {};
\draw  (17.9,6.5) circle (0.125) [fill,circle] {};

\draw (1.05,2.8) -- (1.45,2.8) ;\draw (1.05,2.8) -- (1.05,3) ;\draw (1.45,2.8) -- (1.45,3) ;
\draw[dashed] (0.65,2.8) -- (1.05,2.8) ;\draw[dashed] (0.65,2.8) -- (0.65,3) ;
\draw[dashed] (1.45,2.8) -- (1.85,2.8) ;\draw[dashed] (1.85,2.8) -- (1.85,3) ;
\draw  (1.25,3) circle (0.125) [fill,circle] {};

\draw (3.75,2.8) -- (4.15,2.8) ;\draw (3.75,2.8) -- (3.75,3) ;\draw (4.15,2.8) -- (4.15,3) ;
\draw[dashed] (3.35,2.8) -- (3.75,2.8) ;\draw[dashed] (3.35,2.8) -- (3.35,3) ;
\draw[dashed] (4.15,2.8) -- (4.55,2.8) ;\draw[dashed] (4.55,2.8) -- (4.55,3) ;
\draw  (4.35,3) circle (0.125) [fill,circle] {};

\draw (6.9,2.8) -- (7.3,2.8) ;\draw (6.9,2.8) -- (6.9,3) ;\draw (7.3,2.8) -- (7.3,3) ;
\draw[dashed] (6.5,2.8) -- (6.9,2.8) ;\draw[dashed] (6.5,2.8) -- (6.5,3) ;
\draw[dashed] (7.3,2.8) -- (7.7,2.8) ;\draw[dashed] (7.7,2.8) -- (7.7,3) ;
\draw  (7.1,3) circle (0.125) [fill,circle] {};
\draw  (7.5,3) circle (0.125) [fill,circle] {};

\draw (12.7,2.8) -- (13.1,2.8) ;\draw (12.7,2.8) -- (12.7,3) ;\draw (13.1,2.8) -- (13.1,3) ;
\draw[dashed] (12.3,2.8) -- (12.7,2.8) ;\draw[dashed] (12.3,2.8) -- (12.3,3) ;
\draw[dashed] (13.1,2.8) -- (13.5,2.8) ;\draw[dashed] (13.5,2.8) -- (13.5,3) ;
\draw  (12.5,3) circle (0.125) [fill,circle] {};
\draw  (13.3,3) circle (0.125) [fill,circle] {};

\draw (15.45,2.8) -- (15.85,2.8) ;\draw (15.45,2.8) -- (15.45,3) ;\draw (15.85,2.8) -- (15.85,3) ;
\draw[dashed] (15.05,2.8) -- (15.45,2.8) ;\draw[dashed] (15.05,2.8) -- (15.05,3) ;
\draw[dashed] (15.85,2.8) -- (16.25,2.8) ;\draw[dashed] (16.25,2.8) -- (16.25,3) ;
\draw  (15.25,3) circle (0.125) [fill,circle] {};
\draw  (15.65,3) circle (0.125) [fill,circle] {};

\draw (18.55,2.8) -- (18.95,2.8) ;\draw (18.55,2.8) -- (18.55,3) ;\draw (18.95,2.8) -- (18.95,3) ;
\draw[dashed] (18.15,2.8) -- (18.55,2.8) ;\draw[dashed] (18.15,2.8) -- (18.15,3) ;
\draw[dashed] (18.95,2.8) -- (19.35,2.8) ;\draw[dashed] (19.35,2.8) -- (19.35,3) ;
\draw  (18.35,3) circle (0.125) [fill,circle] {};
\draw  (18.75,3) circle (0.125) [fill,circle] {};
\draw  (19.15,3) circle (0.125) [fill,circle] {};

\draw (1.05,-0.7) -- (1.45,-0.7) ;\draw (1.05,-0.7) -- (1.05,-0.5) ;\draw (1.45,-0.7) -- (1.45,-0.5) ;
\draw[dashed] (0.65,-0.7) -- (1.05,-0.7) ;\draw[dashed] (0.65,-0.7) -- (0.65,-0.5) ;
\draw[dashed] (1.45,-0.7) -- (1.85,-0.7) ;\draw[dashed] (1.85,-0.7) -- (1.85,-0.5) ;
\draw  (1.25,-0.5) circle (0.125) [fill,circle] {};

\draw (3.75,-0.7) -- (4.15,-0.7) ;\draw (3.75,-0.7) -- (3.75,-0.5) ;\draw (4.15,-0.7) -- (4.15,-0.5) ;
\draw[dashed] (3.35,-0.7) -- (3.75,-0.7) ;\draw[dashed] (3.35,-0.7) -- (3.35,-0.5) ;
\draw[dashed] (4.15,-0.7) -- (4.55,-0.7) ;\draw[dashed] (4.55,-0.7) -- (4.55,-0.5) ;
\draw  (4.35,-0.5) circle (0.125) [fill,circle] {};

\draw (6.9,-0.7) -- (7.3,-0.7) ;\draw (6.9,-0.7) -- (6.9,-0.5) ;\draw (7.3,-0.7) -- (7.3,-0.5) ;
\draw[dashed] (6.5,-0.7) -- (6.9,-0.7) ;\draw[dashed] (6.5,-0.7) -- (6.5,-0.5) ;
\draw[dashed] (7.3,-0.7) -- (7.7,-0.7) ;\draw[dashed] (7.7,-0.7) -- (7.7,-0.5) ;
\draw  (7.1,-0.5) circle (0.125) [fill,circle] {};
\draw  (7.5,-0.5) circle (0.125) [fill,circle] {};

\draw (12.7,-0.7) -- (13.1,-0.7) ;\draw (12.7,-0.7) -- (12.7,-0.5) ;\draw (13.1,-0.7) -- (13.1,-0.5) ;
\draw[dashed] (12.3,-0.7) -- (12.7,-0.7) ;\draw[dashed] (12.3,-0.7) -- (12.3,-0.5) ;
\draw[dashed] (13.1,-0.7) -- (13.5,-0.7) ;\draw[dashed] (13.5,-0.7) -- (13.5,-0.5) ;
\draw  (12.5,-0.5) circle (0.125) [fill,circle] {};
\draw  (13.3,-0.5) circle (0.125) [fill,circle] {};

\draw (15.45,-0.7) -- (15.85,-0.7) ;\draw (15.45,-0.7) -- (15.45,-0.5) ;\draw (15.85,-0.7) -- (15.85,-0.5) ;
\draw[dashed] (15.05,-0.7) -- (15.45,-0.7) ;\draw[dashed] (15.05,-0.7) -- (15.05,-0.5) ;
\draw[dashed] (15.85,-0.7) -- (16.25,-0.7) ;\draw[dashed] (16.25,-0.7) -- (16.25,-0.5) ;
\draw  (15.25,-0.5) circle (0.125) [fill,circle] {};
\draw  (15.65,-0.5) circle (0.125) [fill,circle] {};

\draw (18.55,-0.7) -- (18.95,-0.7) ;\draw (18.55,-0.7) -- (18.55,-0.5) ;\draw (18.95,-0.7) -- (18.95,-0.5) ;
\draw[dashed] (18.15,-0.7) -- (18.55,-0.7) ;\draw[dashed] (18.15,-0.7) -- (18.15,-0.5) ;
\draw[dashed] (18.95,-0.7) -- (19.35,-0.7) ;\draw[dashed] (19.35,-0.7) -- (19.35,-0.5) ;
\draw  (18.35,-0.5) circle (0.125) [fill,circle] {};
\draw  (18.75,-0.5) circle (0.125) [fill,circle] {};
\draw  (19.15,-0.5) circle (0.125) [fill,circle] {};

\draw (6.9,-4.2) -- (7.3,-4.2) ;\draw (6.9,-4.2) -- (6.9,-4) ;\draw (7.3,-4.2) -- (7.3,-4) ;

\draw (12.7,-4.2) -- (13.1,-4.2) ;\draw (12.7,-4.2) -- (12.7,-4) ;\draw (13.1,-4.2) -- (13.1,-4) ;
\draw  (12.9,-4) circle (0.125) [fill,circle] {};

\node at (5,8.25) [] {$\frac{ax}{(ax+1)(bx+1)}$} ;
\node at (8.25,8.25) [] {$\frac{abx^2}{(ax+1)(bx+1)}$} ;
\node at (11.75,8.25) [] {$\frac{1}{(ax+1)(bx+1)}$} ;
\node at (15,8.25) [] {$\frac{bx}{(ax+1)(bx+1)}$} ;

\draw[thick] (9.5,9.75) -- (6.5,8.55); \draw[->,thick] (4.5,7.75) -- (2.5,6.95);
\draw[thick] (9.8,9.5) -- (8.8,8.5);  \draw[->,thick] (8,7.7) -- (7.25,6.95);
\draw[thick] (10.2,9.5) -- (11.2,8.5);  \draw[->,thick] (12,7.7) -- (12.75,6.95);
\draw[thick] (10.5,9.75) -- (13.5,8.55); \draw[->,thick] (15.5,7.75) -- (17.5,6.95);

\node at (1.8,4.75) [] {$xz_1$} ;
\node at (3.8,4.75) [] {$1-xz_1$} ;
\node at (7.1,4.75) [] {$1$} ;
\node at (9.4,4.75) [] {$(1-xz_0)(1-xz_1)$} ;
\node at (12.9,4.75) [] {$xz_0(1-xz_1)$} ;
\node at (14.8,4.75) [] {$xz_1$} ;
\node at (18.25,4.75) [] {$1$} ;

\draw[thick] (2.3,6.1) -- (1.8,5.1); \draw[->,thick] (1.5,4.5) -- (1.05,3.6);
\draw[thick] (2.7,6.1) -- (3.2,5.1); \draw[->,thick] (3.5,4.5) -- (3.95,3.6);
\draw[thick] (7.1,6.1) -- (7.1,5.2); \draw[->,thick] (7.1,4.4) -- (7.1,3.6);
\draw[thick] (12.9,6.1) -- (12.9,5.2); \draw[->,thick] (12.9,4.4) -- (12.9,3.6);
\draw[thick] (12.3,6.1) -- (10.8,5.1); \draw[->,thick] (9.9,4.5) -- (8.55,3.6);
\draw[thick] (13.5,6.1) -- (14.5,5.1); \draw[->,thick] (15.1,4.5) -- (16,3.6);
\draw[thick] (17.5,6.1) -- (18,5.1); \draw[->,thick] (18.35,4.4) -- (18.75,3.6);

\node at (1.25,1.25) [] {$\frac{x}{z_1}$} ;
\node at (2.5,1.25) [] {$1-\frac{x}{z_1}$} ;
\node at (3.95,1.25) [] {$1$} ;
\node at (7.1,1.25) [] {$1$} ;
\node at (10,1.25) [] {$1-\frac{x}{z_0}$} ;
\node at (12.9,1.25) [] {$\frac{x}{z_0}$} ;
\node at (14.3,1.25) [] {$1-\frac{x}{z_1}$} ;
\node at (15.65,1.25) [] {$\frac{x}{z_1}$} ;
\node at (18.75,1.25) [] {$1$} ;

\draw[thick] (1.25,2.6) -- (1.25,1.7); \draw[->,thick] (1.25,0.9) -- (1.25,0.1);
\draw[thick] (1.85,2.6) -- (2.35,1.6); \draw[->,thick] (2.65,1) -- (3.1,0.1);
\draw[thick] (3.95,2.6) -- (3.95,1.7); \draw[->,thick] (3.95,0.9) -- (3.95,0.1);
\draw[thick] (7.1,2.6) -- (7.1,1.7); \draw[->,thick] (7.1,0.9) -- (7.1,0.1);
\draw[thick] (12.9,2.6) -- (12.9,1.7); \draw[->,thick] (12.9,0.9) -- (12.9,0.1);
\draw[thick] (15.65,2.6) -- (15.65,1.7); \draw[->,thick] (15.65,0.9) -- (15.65,0.1);
\draw[thick] (18.75,2.6) -- (18.75,1.7); \draw[->,thick] (18.75,0.9) -- (18.75,0.1);
\draw[thick] (12.3,2.6) -- (10.8,1.6); \draw[->,thick] (9.9,1) -- (8.55,0.1);
\draw[thick] (15,2.6) -- (14.5,1.6); \draw[->,thick] (14.2,1) -- (13.75,0.1);

\draw[->,thick] (3.95,-1) -- (6.7,-3.9);
\draw[->,thick] (12.9,-1) -- (7.5,-3.9);
\draw[->,thick] (15.65,-1) -- (13.1,-3.7);
\draw[->,thick] (18.75,-1) -- (13.3,-3.9);
\draw[->,thick] (7.1,-1) -- (12.7,-3.7);
\draw[->,thick] (1.25,-1) -- (12.5,-3.9);
 \end{tikzpicture}
 \caption{An example of sequential update corresponding to the Markov matrix $t(x|\bar z)$. The first line is
 the configuration at time $t$. The second line represents the possible configurations after adding the two supplementary sites 
 corresponding to the boundaries.
 The third line corresponds to the intermediate configurations after the updates of the particles.
 The fourth line represents  the possible configurations after the updates of the holes.
 The label of the arrows provides the rate of the corresponding change of configurations. The last line represents the final
 configurations at time $t+1$ after the summation step.}
 \label{seq.open}
 \end{center}
\end{figure}
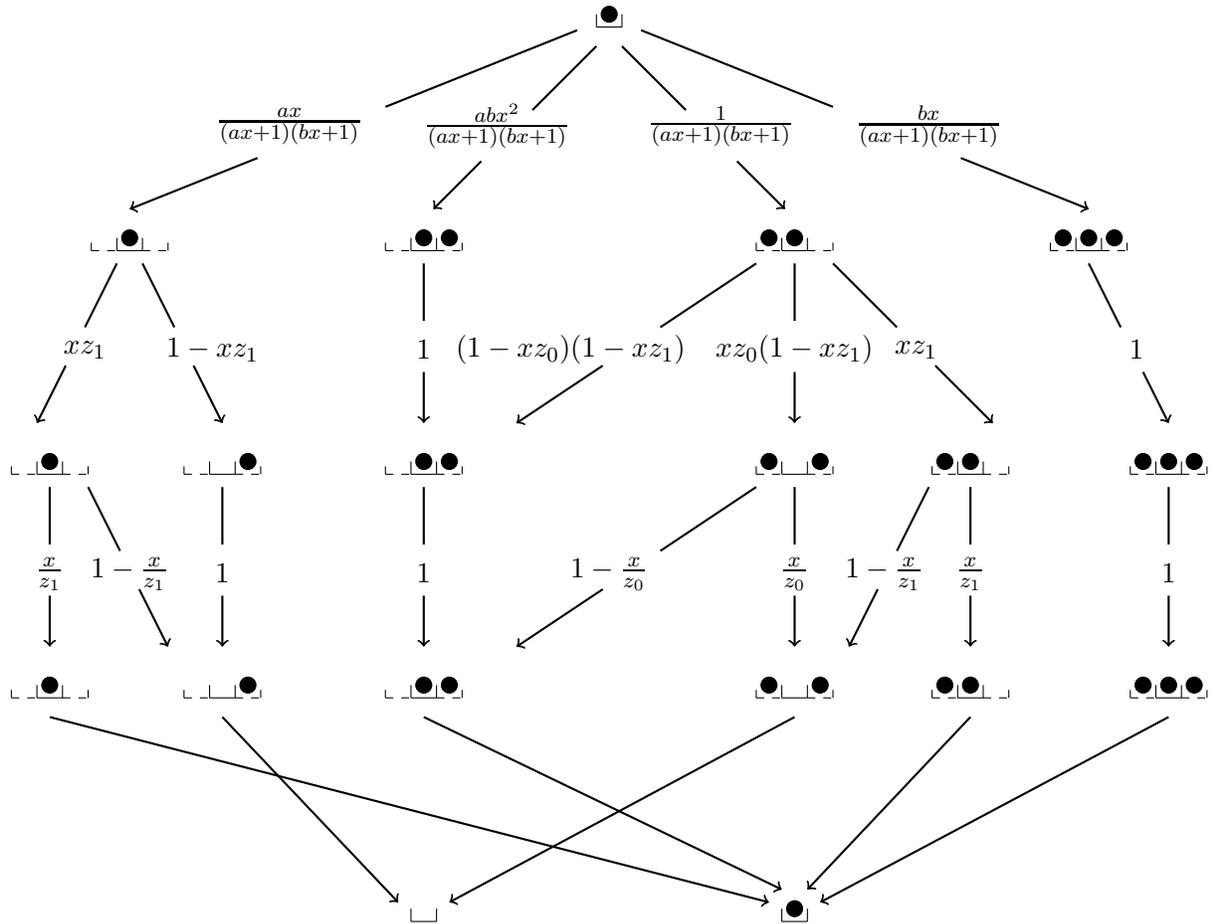

 The transfer matrix $t(x|\bar z)$ defines a discrete time Markov process
 on the finite size lattice with open boundaries. The corresponding
 stochastic dynamics  can be described 
 explicitly using a sequential update: 
starting from a given configuration at time $t$, the configuration at time $t+1$ is obtained  by the  following  stochastic  rules
\begin{itemize}
 \item \textbf{Initialisation:} 
 \begin{itemize}
 \item The left boundary is replaced by an additional site (the site $0$ with inhomogeneity parameter 1) occupied by a particle
 with probability $\frac{1}{ax+1}$ and unoccupied with probability $\frac{ax}{ax+1}$.
 \item The right boundary is replaced by an additional site (the site $L+1$ with inhomogeneity parameter 1) occupied by a particle
 with probability $\frac{bx}{bx+1}$ and unoccupied with probability $\frac{1}{bx+1}$.
 \end{itemize}
 \item \textbf{Particle update:} starting from right to left (from site $L$ to site $0$), a particle at site $i$ can jump to the right on the site $i+1$ (provided 
 that the site is empty) with probability $1-xz_i$ and stay at the same place with probability $xz_i$.
 \item \textbf{Hole update:} once arrived at site $0$, we go the other way starting from left to right (from site $1$ to site $L+1$): a hole at site $i$ can jump
 to the left on the site $i-1$ (provided that the site is occupied) with probability $1-\frac{x}{z_{i-1}}$ and stay at the same place with 
 probability $\frac{x}{z_{i-1}}$.
 \item \textbf{Summation:} Then we have to drop the additional sites $0$ and $L+1$ and  to sum the weights corresponding to 
 the same final configuration.
\end{itemize}
An example of such sequential update is given in fig. \ref{seq.open} for $L=1$. From this figure we can compute the transition rates

\begin{eqnarray}
 \langle 0|t(x|\bar z)|1\rangle & = & \frac{ax}{(ax+1)(bx+1)}\left[xz_1\left(1-\frac{x}{z_1}\right)+1-xz_1\right] \nonu
& & +\frac{1}{(ax+1)(bx+1)}\left[(1-xz_1)x^2+xz_1\left(1-\frac{x}{z_1}\right)\right] \\
& = & (1-x^2)\frac{x(a+z_1)}{(ax+1)(bx+1)},\label{0t1}\\
 \langle 1|t(x|\bar z)|1\rangle & = & \frac{x(x+b)}{bx+1} +\frac{(1-x^2)(1-xz_1)}{(ax+1)(bx+1)}.
\end{eqnarray}

 The equation  \eqref{0t1} is, of course, identical to
  the expression \eqref{weight_example_open}
 derived using the graphical representation in fig.~\ref{fig:open-L=1}.

\subsection{Stationary state and observables}
\label{subsec:StatOpen}

 We now determine  the stationary state of the inhomogeneous Markov chain,
 by using a  matrix product representation, which will be constructed
 following a method developed  in \cite{CRV}. We shall then compute some
 physical characteristics of the steady-state.

\subsubsection{Matrix Product Ansatz}

 The stationary state is expressed as a  matrix product.
Indeed,  the probability of
the configuration $\mathcal{C}=(\tau_1,\dots,\tau_L)$ can be written as
\begin{equation}
 \mathcal{S}(\tau_1,\dots,\tau_L)=\frac{1}{Z_L}\llangle W| \prod_{i=1}^{L} ((1-\tau_i)E(z_i)+\tau_iD(z_i)) |V \rrangle \;,
\end{equation}
where the normalization factor is 
$Z_L=\llangle W|C(z_1)\dots C(z_L) |V\rrangle$, with $C(z)=E(z)+D(z)$.
 The algebraic elements $E(z)$ and $D(z)$ can be gathered in a vector 
\begin{equation}
 A(z)=\left(
 \begin{array}{c}
 E(z)\\ D(z)
 \end{array}
\right). 
\end{equation}
Using this vector, the stationary state can be written in a more compact way: 
\begin{equation} \label{steady_open}
 \steady = \frac{1}{Z_L}\llangle W|A_1(z_1)A_2(z_2)\dots A_L(z_L)|V\rrangle \;.
\end{equation}

 We now use  the formalism first introduced in \cite{Sasamoto2} and
 developed  in \cite{CRV,CMRV}
 to determine the algebra
 generated  by the matrix elements of  $A(z)$ and by the boundary vectors.
The quadratic  algebra  satisfied  in the bulk  by 
  $E$ and $D$ will be obtained using  the  Zamolodchikov-Faddeev relations; 
  the boundary Ghoshal-Zamolodchikov relations
   show how  the vectors $ \llangle W |$ and $ |V \rrangle$ are coupled to  $E$ and $D$. 

The vector $A(z)$ allows us to write easily the exchange relations
 between $E(z)$ and $D(z)$:
\begin{equation} \label{eq:ZF}
 R_{12}(z_1/z_2)A_1(z_1)A_2(z_2)=A_2(z_2)A_1(z_1).
\end{equation}
This {\it Zamolodchikov-Faddeev relation}  \cite{ZF}
 defines an associative algebra because the $R$-matrix obeys the 
Yang-Baxter equation.
Written explicitly it gives four relations
\begin{equation} \label{ZF_components}
 \left\{
 \begin{array}{l}
 \left[E(z_1),E(z_2)\right]=0\\[1.ex] 
 \left[D(z_1),D(z_2)\right]=0\\[1.ex]
 \frac{z_1}{z_2}D(z_1)E(z_2)=D(z_2)E(z_1)\\[1.ex]
 D(z_1)E(z_2)=\frac{z_2}{z_2-z_1}\left[ E(z_2)D(z_1)-E(z_1)D(z_2) \right]
 \end{array}
 \right.
\end{equation}
Let us remark that the third relation is implied by the fourth.

The relations between the boundary vectors $\llangle W|$, $|V\rrangle$ and the algebraic elements $E(z)$, $D(z)$ 
are given by the {\it   Ghoshal-Zamolodchikov relations}
\begin{equation} \label{eq:GZ}
 \llangle W | K(z)A(1/z)=\llangle W| A(z), \qquad \bar K(z)A(1/z) |V \rrangle = A(z) |V \rrangle  \;.
\end{equation}
These boundary relations are consistent with the bulk exchange relations thanks to the reflection equation.
Remark that the boundary vector $|V \rrangle$ (resp. 
$\llangle W|$) in fact defines a class of representations (resp. dual representations)  for the ZF algebra, where $|V \rrangle$ (resp. 
$\llangle W|$) exists and is non-vanishing. We prove below that these classes are not empty by providing an explicit (dual) representation that belongs to them.

Written explicitly we obtain  four relations
\begin{equation} \label{GZ_components}
 \left\{
 \begin{array}{l}
 \llangle W|E(z)=\llangle W|\frac{(a+z)z}{za+1}E(1/z) \\[1.ex]
 \llangle W|E(z)=\frac{z(a+z)}{z^2-1}\llangle W|(D(1/z)-D(z) )
 \end{array}
 \right.
 \text{ and  }
  \left\{
 \begin{array}{l}
 D(z)|V\rrangle= \frac{zb+1}{z(b+z)}D(1/z)|V\rrangle \\[1.ex]
 D(z)|V\rrangle=\frac{bz+1}{z^2-1}(E(z)-E(1/z) )|V\rrangle
 \end{array}
 \right.
\end{equation}
Let us remark that for each boundary, the first relation is implied by the second one.

We  now prove that the vector $\steady$ is the stationary state of the transfer matrix. Following what was done in \cite{CRV},
one picks up the vector $A_i(z_i)$ and moves it completely to the right using $L-i$ times the Zamolodchikov-Faddeev relation \eqref{eq:ZF}.
Next, the vector $A_i(z_i)$ can be reflected against the right boundary vector $|V\rrangle$ using the Ghoshal-Zamolodchikov relation
\eqref{eq:GZ}. After that, one can push $A_i(1/z_i)$ to the left through all the $A_j(z_j)$ using $L-1$ times \eqref{eq:ZF}.
Then, one can reflect $A_i(1/z_i)$ against the left boundary vector $\llangle W|$ using \eqref{eq:GZ}. Finally, one can move back $A_i(z_i)$ 
to its initial position to complete the cycle. Thus,  we are left with:
\begin{equation}
 t(z_i|\bar z)\steady = \steady \;.
\end{equation}
If we do the previous cycle the other way round we get
\begin{equation}
 t(1/z_i|\bar z)\steady = \steady \;.
\end{equation}
We show below, using the graphical representation, that the numerators
 of the entries of $t(x|\bar z)$ are polynomials of degree
less than $L+2$. Since the equation $t(x|\bar z)\steady = \steady$ is satisfied for $2L$ different values of $x$ ($x=z_1,1/z_1,\dots,z_L,1/z_L$),
we get through interpolation arguments that for $L\geq 3$, $t(x|\bar z)\steady = \steady$ for all $x$. For $L<5$ we checked by computer  that this equation
is satisfied.

\begin{figure}[h]
\begin{center}
 \begin{tikzpicture}[scale=1]
\draw[->,ultra thick] (0,0) -- (0.5,0); \draw[ultra thick] (0.5,0) -- (2,0);
\draw[->,ultra thick] (1,-1) -- (1,-0.5); \draw[ultra thick] (1,-0.5) -- (1,1);
\draw[->,ultra thick] (4,0) -- (4.5,0); \draw[ultra thick] (4.5,0) -- (4.75,0); \draw[ultra thick] (5.25,0) -- (6,0);
\draw[->,ultra thick] (5,-1) -- (5,-0.5); \draw[ultra thick] (5,-0.5) -- (5,-0.25); \draw[ultra thick] (5,0.25) -- (5,1);
\draw[ultra thick] (4.75,0) arc (-90:0:0.25); \draw[ultra thick] (5.25,0) arc (90:180:0.25);
\node at (3,0) [] {$\equiv$} ;
 \end{tikzpicture}
 \end{center}
 \caption{Vertex with non-intersecting thick lines.\label{identification:vertex}}
\end{figure}
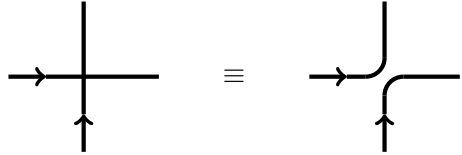

\paragraph{Degree of the numerator of  $t(x|\bar z)$:}
We use the graphical representation given in figures \ref{matrix_elements}, \ref{fig:Kmat} and \ref{fig:Ktilde}, interpreting the last vertex of fig. \ref{matrix_elements} as in fig. \ref{identification:vertex}.
One can see that any matrix element $\langle \cC'|t(x|\bar z)|\cC\rangle$ can be decomposed into continuous thick lines that do not intersect, as illustrated in the  figure \ref{thicklines}.
\begin{figure}[h]
\begin{center}
 \begin{tikzpicture}[scale=0.7]
\draw (0,-2) -- (0,2) ;
\draw (12,-2) -- (12,2) ;
\foreach \i in {-1.6,-1.2,...,2.1}
{\draw (-0.4,\i-0.4) -- (0,\i) ; \draw (12+0.4,\i-0.4) -- (12,\i) ;}
\draw[->,dashed] (2,-2) -- (1,-1); \draw[dashed] (1,-1) -- (0,0); \draw[->,dashed] (0,0) -- (1,1); \draw[dashed] (1,1) -- (2,2);
\draw[->,ultra thick] (10,2) -- (11,1); \draw[ultra thick] (11,1) -- (12,0); \draw[->,dashed] (12,0) -- (11,-1); \draw[dashed] (11,-1) -- (10,-2);
\draw[dashed] (2,2) -- (3,2); \draw[ultra thick] (3,2) -- (4.75,2); \draw[ultra thick] (5.25,2) -- (8.75,2); \draw[ultra thick] (9.25,2) -- (10,2);
\draw[dashed] (2,-2) -- (3,-2); \draw[dashed] (7,-2) -- (10,-2); \draw[ultra thick] (3,-2) -- (4.75,-2); \draw[ultra thick] (5.25,-2) -- (7,-2);
\draw[->,dashed] (3,-3) -- (3,-2.5); \draw[dashed] (3,-2.5) -- (3,-2); \draw[->,dashed] (3,2) -- (3,2.5); \draw[dashed] (3,2.5) -- (3,3); \draw[ultra thick] (3,-2) -- (3,2);
\draw[->,ultra thick] (5,-3) -- (5,-2.5); \draw[ultra thick] (5,-2.5) -- (5,-2.25); \draw[ultra thick] (5,-1.75) -- (5,1.75); \draw[->,ultra thick] (5,2.25) -- (5,2.5); \draw[ultra thick] (5,2.5) -- (5,3);
\draw[->,dashed] (7,-2) -- (7,2.5); \draw[dashed] (7,2.5) -- (7,3); \draw[->,ultra thick] (7,-3) -- (7,-2.5); \draw[ultra thick] (7,-2.5) -- (7,-2);
\draw[->,ultra thick] (9,-3) -- (9,-2.5); \draw[ultra thick] (9,-2.5) -- (9,1.75); \draw[->,ultra thick] (9,2.25) -- (9,2.5); \draw[ultra thick] (9,2.5) -- (9,3);
\draw[ultra thick] (5,-2.25) arc (0:90:0.25); \draw[ultra thick] (5,-1.75) arc (180:270:0.25);
\draw[ultra thick] (4.75,2) arc (270:360:0.25); \draw[ultra thick] (5.25,2) arc (90:180:0.25);
\draw[ultra thick] (8.75,2) arc (270:360:0.25); \draw[ultra thick] (9.25,2) arc (90:180:0.25);
 \end{tikzpicture}
 \caption{Decomposition of a matrix element into thick continuous lines.\label{thicklines}}
 \end{center}
\end{figure}

The vertices that do not enter a thick line have degree 0 in $x$. Thus, the total degree of the matrix element is the sum of the degree of each thick line, plus the degree coming from the boundaries that we look at separately. We first consider lines that are in the 'bulk' (that do not include a boundary). Looking at the weights of the vertices in fig. \ref{matrix_elements}, one can see that the incoming part of a line always carry a degree 1. Moreover,
a line corresponding to a degree $n$ must cross at least $n$ exit lines, the first on the left being always of degree 0. This implies that the bulk part is at most of degree $L$. Figure  \ref{fig:Ktilde} shows that the right boundary does not change this counting. The left boundary can add at most a degree 2, as it can be seen on figure  \ref{fig:Kmat}. Altogether, this
leads to a total degree $L+2$.

\paragraph{Explicit representation.}
We now  construct an explicit representation of the algebraic elements $E(z)$ and $D(z)$. 
 We  introduce 
the shift  operators $\epsilon$ and $\delta$ such that $\delta \epsilon=1$, which 
  are well known in the context of the continuous time open
  TASEP since they enter the construction of the original matrix ansatz \cite{DEHP}.
We define  
\begin{equation}
 \widetilde{E}(z)=z+\epsilon \,\,\,\, \hbox{ and } \,\,\,\,
  \widetilde{D}(z)=1/z+\delta \, ,
\label{ReprExpl}
\end{equation}
and the boundary vectors such that 
\begin{equation}\label{eq:bound-ed}
\llangle\wt W|\, \epsilon = a\,\llangle\wt W|  \,\,\,\, \hbox{ and } \,\,\,\,
\delta |\wt V\rrangle=b\,|\wt V\rrangle \, .
\end{equation}
It is  readily verified that   the 
Zamolodchikov-Faddeev relation \eqref{eq:ZF} and 
the Ghoshal-Zamo\-lo\-dchikov relations \eqref{eq:GZ} are satisfied. 
From the results of \cite{DEHP} giving explicit forms for $\epsilon$, $\delta$, $\llangle\wt W|$ and $|\wt V\rrangle$, we deduce 
that $\llangle\wt W|$ and $|\wt V\rrangle$ can be chosen such that $\llangle\wt W|\wt V\rrangle\neq 0$ which guarantees the non-vanishing of $\steady$.

\subsubsection{Steady-state probabilities} \label{sec_steady}

The matrix Ansatz allows us to calculate the 
 stationary probability  of any given configuration. 
Let $1\leq j_1<\dots<j_r\leq L$ be integers and $\mathcal{C}(j_1,\dots,j_r)$ be the configuration
$(\tau_1,\dots,\tau_L)$ with $\tau_i=1$ if $i=j_k$ and $\tau_i=0$ otherwise.
We define the (non-normalized)
 weight of the word with $D(j_1),\dots, D(j_r)$ at positions $j_1,\dots,j_r$ as 
\begin{eqnarray}
 W_L(j_1,\dots,j_r) & = & Z_L \times \mathcal{S}(\mathcal{C}(j_1,\dots,j_r)) \\
 & = & \llangle W | \dots D(z_{j_1}) \dots D(z_{j_2}) \dots D(z_{j_r}) \dots |V\rrangle,
\end{eqnarray}
where the dots stand for $E(z_i)$ operators.

The same quantity can be calculated using  the explicit representation \eqref{ReprExpl}:
\begin{equation}
 \widetilde{W}_L(j_1,\dots,j_r)= \llangle\wt W| \dots \left(\delta +\frac{1}{z_{j_1}}\right) \dots \left(\delta +\frac{1}{z_{j_2}}\right)
 \dots \left(\delta +\frac{1}{z_{j_r}}\right) \dots |\wt V\rrangle.
\end{equation}
 The weight  $W_L(j_1,\dots,j_r)$ computed from  the relations
  \eqref{ZF_components} and  \eqref{GZ_components}  is  proportional to   the weight 
 $\widetilde{W}_L(j_1,\dots,j_r)$ computed using the explicit representation 
 (thanks to  the unicity of the steady-state guaranteed by the Perron-Frobenius theorem). Thus, we have 
\begin{equation} \label{equivalence_rep_explicite}
 W_L(j_1,\dots,j_r)=f(z_1,\dots,z_L)\widetilde{W}_L(j_1,\dots,j_r).
\end{equation}
 The multiplicative coefficient is obtained by comparing the weights of the empty
 configuration:
\begin{equation}
 f(z_1,\dots,z_L)=
 \frac{\llangle W| E(z_1)\dots E(z_L) |V\rrangle}
{ (a+z_1)\dots (a+z_L)\llangle W|V\rrangle} \, ; 
\end{equation}
$f(z_1,\dots,z_L)$ is symmetric under the permutation of the $z_i$
 thanks to \eqref{ZF_components} but also under the transformation
$z_i \mapsto 1/z_i$ thanks to \eqref{GZ_components}.
{The group generated by these transformations is denoted by $BC_L$ in reference to the Weyl group
of the root system of the Lie algebra $sp(2L)$.}

The expression of  $\widetilde{W}_L(j_1,\dots,j_r)$ is given by
\begin{eqnarray}
 \widetilde{W}_L(j_1,\dots,j_r) & = &  \frac{z_1\dots z_L}{z_{j_1} \dots z_{j_r}}
 \sum_{p=0}^{r}\,b^{p}\sum_{q_{r-p}=j_{r-p}}^{L}\frac1{z_{q_{r-p}}}\,\sum_{q_{r-p-1}=j_{r-p-1}}^{q_{r-p}-1}\frac1{z_{q_{r-p-1}}} \dots 
 \sum_{q_1=j_1}^{q_2-1}\frac{1}{z_{q_1}} \nonu
 & & \times \prod_{l_0=1}^{j_1-1}\left(1+\frac{a}{z_{l_0}}\right)\prod_{l_1=q_1+1}^{j_2-1}\left(1+\frac{a}{z_{l_1}}\right)
 \dots \prod_{l_{r-p}=q_{r-p}+1}^{j_{r-p+1}-1}\left(1+\frac{a}{z_{l_{r-p}}}\right)
\end{eqnarray}
By convention when $p=r$ in the first sum, there is no summation over the $q_i$ and the formula  reduces to
$b^r\prod_{l_0=1}^{j_1-1}\left(1+\frac{a}{z_{l_0}}\right)$. We also set $j_{r+1}=L+1$ in the last product when $p=0$. 
The proof of this formula is obtained by induction on the size $L$, using the identity
\begin{eqnarray}
 \widetilde D(z_{j_r}) \widetilde E(z_{j_r+1})&=& \left( \frac{1}{z_{j_r}}+\delta \right)\left(z_{j_r+1}+\epsilon \right)=
 \frac{1}{z_{j_r}}\left(z_{j_r+1}+\epsilon \right)+z_{j_r+1}\left(\frac{1}{z_{j_r+1}}+\delta \right)\\
& =& \frac1{z_{j_r}} \widetilde E(z_{j_r+1}) + z_{j_r+1} \widetilde D(z_{j_r+1}) \, . 
\end{eqnarray}

\vskip 0.3cm
Another important step is to compute the normalization factor
 of the probability distribution.
 We define  $C(z)=E(z)+D(z)$ and
 $\widetilde{C}(z)=\widetilde{E}(z)+\widetilde{D}(z)$.
The normalization factors are  thus given by 
 $Z_L(z_1,\dots,z_L)=\llangle W|C(z_1)\dots C(z_L)|V\rrangle$, and
$\widetilde{Z}_L(z_1,\dots,z_L)=\llangle \wt W|\widetilde{C}(z_1)\dots \widetilde{C}(z_L)|\wt V\rrangle$ for the explicit representation.
Thanks to the property \eqref{equivalence_rep_explicite}, we have 
\begin{equation}
 Z_L(z_1,\dots,z_L)=f(z_1,\dots,z_L)\widetilde{Z}_L(z_1,\dots,z_L).
\end{equation}
$Z_L$ and $\widetilde{Z}_L$ are symmetric under $BC_L$.

 Our goal is to get an analytic expression for the 
normalization factor.
{For this purpose, for any sequence of complex numbers  $\wa=(u_1,u_2,...)$, we define the shifted product by
\begin{equation}
(z|\wa)^{k}=\begin{cases} (z-u_1)(z-u_2)\cdots(z-u_k)\,,\quad k>0, \\ 1 \mb{if} k=0,\end{cases}
\end{equation}}
We have the result (see proof in appendix \ref{proof_normalzation}):
\begin{equation}\label{schurCshifte}
 \widetilde{Z}_L(z_1,\dots,z_L)= \frac{\det \left((z_j|\va)^{L+2-i}-(1/z_j|\va)^{L+2-i} \right)_{i,j}}{\det \left(z_j^{L+1-i}-(1/z_j)^{L+1-i} \right)_{i,j}}
\quad \mbox{with}\quad \va=(-a,-b,0,\dots,0).
\end{equation}
When $a,b=0$, we recognize {in the L.H.S. of  equation \eqref{schurCshifte}, the expression of the Schur polynomial of type C associated to the partition $1^L=(1,....,1)$. We remind that}
the Schur polynomial of type C associated with the partition $\lambda=(\lambda_1,\dots,\lambda_L)$ 
with $\lambda_1\geq \dots \lambda_L\geq 0$ is defined by:
\begin{equation}\label{eq:sC}
\widetilde{Z}_L(z_1,\dots,z_L)\Big|_{a=b=0}=s^C_{\lambda}(z_1,\dots,z_L)= \frac{\det \left((z_j)^{L+1-i+\lambda_i}-(1/z_j)^{L+1-i+\lambda_i} \right)_{i,j}}{\det \left(z_j^{L+1-i}-(1/z_j)^{L+1-i} \right)_{i,j}}\;.
\end{equation}
As in the periodic case where the normalization factor is linked to the Schur polynomial 
of type $A$ and to the  representation
of the Lie algebra $sl(L)$ (see \eqref{eq:rsl1} and \eqref{eq:rsl2}), the normalization factor \eqref{eq:sC} is given in terms of the Schur polynomial of type $C$ 
and is associated to representation of the Lie algebra $sp(2L)$. These observations will 
 allow us to use  results of the Lie algebra theory to take the homogeneous 
limit $z_i\rightarrow 1$ (see the  remark below).

When $a,b$ are arbitrary, we need to use some 
 generalizations of the Schur polynomial,  called {\it shifted (or factorial) Schur polynomials}.
A shifted  Schur polynomial 
 of type A, is defined, 
for any sequence of complex numbers $\wa=(u_1,u_2,...)$ \cite{SchurPoly}, as follows:
\bea\label{eq:sA2}
s^A_{\lambda}(z_1,\dots,z_L|\wa) &=& \frac{\det \left((z_j|\wa)^{L-i+\lambda_i} \right)_{i,j}}{\det \left(z_j^{L-i} \right)_{i,j}}\;.
\eea
Similarly, we define the shifted (or factorial) Schur polynomial of type C as\footnote{{It is easy to see that 
$s^C_{\lambda}(z_1,\dots,z_L|\wa)$ is indeed a polynomial in the variables $z_1,\dots,z_L$.}}
\begin{equation}\label{eq:sC2}
s^C_{\lambda}(z_1,\dots,z_L|\wa)= \frac{\det \left((z_j|\wa)^{L+1-i+\lambda_i}-(1/z_j|\wa)^{L+1-i+\lambda_i} \right)_{i,j}}{\det \left(z_j^{L+1-i}-(1/z_j)^{L+1-i} \right)_{i,j}}\;.
\end{equation}
Thus,  the normalization factor \eqref{schurCshifte} is the shifted Schur polynomial
 of type C associated to the partition $1^L$ 
and to the sequence $\va=(-a,-b,0,0,\dots)$.
For this particular partition, the shifted Schur polynomial of type C can be expanded on the usual type C Schur polynomials as (see proof in appendix \ref{app.schur})
\begin{equation}\label{eq:ex}
\widetilde{Z}_L(z_1,\dots,z_L)=s^C_{1^L}(z_1,\dots,z_L|\va)=\sum_{n=1}^{L+1} \frac{a^n-b^n}{a-b}\, s^C_{1^{L+1-n}}(z_1,\dots,z_L).
\end{equation}
\hfill\break 

Using the explicit representation, we can also determine the particle density at site $i$ from the stationary measure:
\begin{eqnarray*}
 n_i(z_1,\dots,z_L) & = & \left(b+\frac{1}{z_i}\right) \frac{\widetilde{Z}_{L-1}(z_1,\dots,z_{i-1},z_{i+1},\dots,z_L)}{\widetilde{Z}_L(z_1,\dots,z_L)} + \\
 (1-ab) & \times & \sum_{k=0}^{L-i-1} 
 \frac{\left. \widetilde{Z}_{i-1+k}(z_1,\dots,z_{i-1},z_{i+1},\dots,z_{i+k})\right|_{b=0}\left. \widetilde{Z}_{L-i-1-k}(z_{i+2+k},\dots,z_L)\right|_{a=0}}{\widetilde{Z}_L(z_1,\dots,z_L)}.
\end{eqnarray*}

\paragraph{Remark:} 
In the homogeneous
limit $z_i \rightarrow 1$,  $ s^C_{1^{L+1-n}}(1,\dots,1)$ is equal to the dimension of  the $sp(2L)$ representation   
associated to the partition $1^{L+1-n}$:
{
\be\label{eq.dimC}
s^C_{1^{L+1-n}}(1,\dots,1)=\frac{n}{L+1}\,
\left( \begin{array}{c}  2L+2\\  L+1-n  \end{array} \right)
=dim\left(\pi^C(1^{L+1-n})\right).
\ee
This expression is obtained through the general formula (see \cite{dico} for instance):
$$
dim\left(\pi^C(\lambda)\right)= \prod_{1\leq i<j\leq L} \left(\frac{\lambda_i-\lambda_j+j-i}{j-i}\ \frac{\lambda_i+\lambda_j+2L+2-j-i}{2L+2-j-i}\right)\ \prod_{i=1}^{ L} \frac{\lambda_i+L+1-i}{L+1-i}.
$$
} 
Then from \eqref{eq:ex}, we get 
\begin{equation}\label{eq.ZL.C}
\widetilde{Z}_L(1,1,...,1) = \frac{g_L(a)-g_L(b)}{a-b}
\mb{with} g_L(x)=\sum_{n=0}^{L} \frac{n+1}{L+1}\,
\left( \begin{array}{c}  2L+2\\  L-n  \end{array} \right)\,x^{n+1}
\end{equation}
in accordance with the results known for continuous time open TASEP \cite{DEHP}
\begin{equation} \label{eq.ZL.Derrida}
Z_L =  \frac{h_L(\frac1{\alpha})-h_L(\frac1{\beta})}{\frac{1}{\alpha}-\frac{1}{\beta}}
\mb{with} h_L(x)=\sum_{p=1}^{L} \frac{p}{2L-p}
\left( \begin{array}{c}  2L-p\\  L  \end{array} \right)\,x^{p+1}.
\end{equation}
{ The equality between these two expressions is ensured by 
the identity
\begin{equation} \label{eq:egaliteZL}
g_L(x-1)=h_L(x)-\frac1L\left(\begin{array}{c} 2L\\L-1\end{array}\right)
\end{equation}
} and recalling that $a=\frac1{\alpha}-1$ and $b=\frac1{\beta}-1$. 

In particular, when $a=b=0$ (i.e. $\alpha=\beta=1$),  $Z_L$
 is equal to the  dimension of the representation $1^L$,  which is  the Catalan number
$ C_{L+1}=\frac{1}{L+2}\left( \begin{array}{c}
                              2L+2\\
                              L+1
                             \end{array} \right).
$

\hfill\break

\subsection{Partition function and Baxter $Q$-operator \label{sec:bly}}

The models  considered  here have non-diagonal boundary matrices (see \eqref{eq:K}) and cannot be solved by
the usual integrability methods. In the last few years, various
   specific techniques have been developed for solving such problems, based on 
 the functional Bethe ansatz \cite{nepo,MurN05,FraGSW11},
the coordinate Bethe ansatz \cite{CR,CRS1,CRS2,S}, the separation of variables \cite{FSW,niccoli2}, 
 the q-Onsager approach \cite{BK} and  the algebraic Bethe ansatz \cite{BCR,BC}.
Recently, a generalization \cite{CYSW1,Nep} of the TQ relations expresses the 
  eigenvalues of   problems  in terms of 
solutions of a new type of Bethe equations, called {\it inhomogeneous Bethe equations.}

In this section, we shall find   an unexpected  relation
 between  the `partition function' $Z_L$ and  the Baxter $Q$ operator
 appearing in the  $TQ$-relations.

In \cite{crampe13}, the Bethe equations corresponding to the eigenvalues and the eigenvectors of the transfer matrix \eqref{eq:transfer_matrix} 
have been computed
using the modified algebraic Bethe ansatz.
In this context, the eigenvalue of \eqref{eq:transfer_matrix} corresponding to the stationary state $\Lambda(x)$ is given by
\begin{equation}\label{eq:L}
 \Lambda(x)=x^{L+1}\frac{b+x}{bx+1} \prod_{k=1}^L \frac{xu_k-1}{u_k-x}
 -\frac{(x^2-1)}{(bx+1)}\prod_{j=1}^L\left[(x-z_j)(x-\frac{1}{z_j})\right] \prod_{k=1}^L \frac{u_k}{u_k-x}\;,
\end{equation}
where $u_k$ are called Bethe roots and are solutions of the following Bethe equations
\begin{equation}\label{eq:beinh}
\prod_{p=1}^L\frac{(u_j-z_p)(u_jz_p-1)}{u_jz_p}=  (u_j+b)\  \prod_{\ato{k=1}{k\neq j}}^{L}
 \left(u_j-\frac{1}{u_k}\right)\ ,\qquad\text{for $j=1,2,\dots,L$}\;.
\end{equation}
Let us stress that here the Bethe roots are the ones corresponding to the steady state. The Bethe roots corresponding to other states obey 
different Bethe equations (see \cite{crampe13}, or \cite{DE} for the homogeneous case).
We  now introduce the following function:
\begin{equation}\label{eq:Q}
 Q(x)=\prod_{k=1}^L\left(\frac{1}{u_k}-\frac{1}{x}\right)\;.
\end{equation}
This function (up to a coefficient $x^L$) is linked to the polynomial $Q$ of Baxter \cite{Baxt-book}: its zeros are the Bethe roots.
The eigenvalue \eqref{eq:L} can be written in terms of the Q function as follows
\begin{equation}\label{eq:LQ}
 \Lambda(x)=\frac{x(b+x)Q(1/x)}{(bx+1)Q(x)}
 -\frac{(x^2-1)}{(bx+1)Q(x)}\prod_{j=1}^L\left[(x-z_j)(\frac{1}{xz_j}-1)\right]\;.
\end{equation}
For the steady-state  eigenvector, we know  that $\Lambda(x)=1$ (see section \ref{subsec:StatOpen}). 
Hence  $Q(x)$  satisfies
\begin{equation}\label{eq:TQ}
x(x+b)Q(1/x)-(1+bx)Q(x)=(x^2-1) \prod_{j=1}^L\left[(x-z_j)(\frac{1}{xz_j}-1)\right]\;.
\end{equation}
This equation, called TQ relation,  allows us to compute explicitly the function Q:
 for a given $L$, it can be shown that equation \eqref{eq:TQ} has a unique solution of the form \eqref{eq:Q}.

For the model studied here,  the function Q and the normalization factor $\widetilde Z(z_1,\dots,z_L)$
are closely related. 
Namely, we get
\begin{eqnarray}
 Q(x)= \widetilde Z(z_1,\dots,z_L)\big|_{a\rightarrow -1/x}&=&\sum_{n=1}^{L+1} \frac{(-1/x)^n-b^n}{-1/x-b}\, s^C_{1^{L+1-n}}(z_1,\dots,z_L)
 \label{eq:QZ1}\\
 &=&\sum_{p=0}^L \left(-\frac{1}{x}\right)^p\  \sum_{n=0}^{L-p} b^n\ \  s^C_{1^{L-n-p}}(z_1,\dots,z_L)\;.\label{eq:QZ2}
\end{eqnarray}
where we have used the explicit form of $\widetilde Z$ given in \eqref{eq:ex}.
To prove this result, we remark that $Q(x)$ given by \eqref{eq:QZ2} has the  form \eqref{eq:Q}
and we show that it satisfies the TQ relation \eqref{eq:TQ}. We  have 
\begin{eqnarray}
 x(x+b)Q(1/x)-(1+bx)Q(x)&=&-x\sum_{n=1}^{L+1}\left((-x)^n-(-1/x)^n\right)s^C_{1^{L+1-n}}(z_1,\dots,z_L)\\
 &=&(x^2-1)s^C_{1^{L}}(z_1,\dots,z_L|x,1/x,0,\dots,0)\label{eq:TQp}\;.
\end{eqnarray}
Finally, we readily check  that $s^C_{1^{L}}(z_1,\dots,z_L|x,1/x,0,\dots,0)$ vanishes at the points $x=z_j$ and $x=1/z_j$
which allows us to conclude that \eqref{eq:TQp} is equal to the L.H.S. of \eqref{eq:TQ}.

 Relation \eqref{eq:QZ1} means that the Bethe roots (zeros of the function $Q$) are linked to the zeros
of the steady-state normalization factor in the complex plane of the transition rate $a$. These zeros 
appeared previously (for the homogeneous  case) in \cite{LY,BDL} as Lee-Yang zeros and allows the generalization of the Lee-Yang theory 
for the phase transition of non equilibrium system. Therefore, relation \eqref{eq:QZ1} expresses
 an unexpected relation between  two  objects arising from very different contexts.

\section{Conclusion}

In this paper, we study discrete time Markov processes with periodic or open boundaries and with inhomogeneities.
The transfer matrices associated with 
quantum spin chains are used to construct the Markov matrix. We provide a graphical representation of these processes and an interpretation in terms of sequential updates.
We also study their stationary states using a matrix ansatz developed in \cite{Sasamoto2,CRV} and express their normalization in terms of Schur polynomials. 
Finally, in the case with open boundaries, 
we find a connection between the Bethe roots and the Lee-Yang zeros.  \\

We believe that numerous generalizations of these works are possible and that connections with other approaches will be fruitful.
For the moment, we studied the models 
on general grounds, without specifying the values of the inhomogeneities.
However, it would be interesting to take particular choices 
inspired by physical problems. In these cases, we may be able to go further in the computations of the physical quantities.
For example, one can choose the inhomogeneities randomly, according to a certain law,  and study the behavior of the density and of the current.

To compute the stationary state, we use a generalization of the matrix ansatz but other methods have been developed to study such problems.
For example, the study of qKZ equations \cite{DIZ,cantini2} seems to be closely related
 to  the approach developed here.
A common point of both methods is the use of different types of 
symmetric polynomials (or their deformation). Let us also mention the combinatorial interpretation of the matrix ansatz developed in \cite{CW1} 
which may be generalized to the case with inhomogeneities. 

The method presented here can be  used for other models:
for example, in a recent work \cite{CMRV}, we classified integrable boundary conditions
 for the open two species exclusion process
generalizing the model introduced in \cite{uch}  (see also \cite{cantini1,CW2}).
Thus, discrete time inhomogeneous two species exclusion processes
 can be constructed and studied following the lines presented here. 
The generalization to $N$-species also seems  possible \cite{Cantini-P}.

Finally, the correspondence between the Bethe roots and the Lee-Yang zeros  deserves
 further investigation. 
We do not know  if it is valid only for the TASEP or if it is a general feature.
Such  connection   may shed light  on the Bethe 
equations and  on the Lee-Yang theory for out-of-equilibrium models \cite{LY,MartinRev}.

\section*{Acknowledgments}
It is a pleasure to thank A. Molev for discussions and references on Schur polynomials. M. Vanicat thanks the Laboratoire Charles Coulomb 
for hospitality during his stay.

\appendix

\section{Normalization factor and Schur polynomials of type C} \label{proof_normalzation}
The normalization factor can be expressed in terms of a shifted Schur polynomial of type C
 \begin{equation}
  \widetilde Z (z_1,\dots,z_L)=s_{1^L}^C(z_1,\dots,z_L|{\bf v})
 \end{equation}
 
 \paragraph{Proof}
Let us show be induction on $L$ the previous formula.
\begin{itemize}
 \item For $L=1$, we have
 \begin{equation*}
 \widetilde Z_1(z_1)=\llangle\wt W | \left(z_1+ \frac{1}{z_1}+ \epsilon + \delta \right)|\wt V\rrangle =z_1+\frac{1}{z_1}+a+b,
 \end{equation*}
 in agreement with
 \begin{equation*}
  s_{1^1}^C(z_1|{\bf v})= \frac{(z_1+a)(z_1+b)-\left(\frac{1}{z_1}+a\right)\left(\frac{1}{z_1}+b\right)}{z_1-\frac{1}{z_1}}=z_1+\frac{1}{z_1}+a+b.
 \end{equation*}
 \item Let $L\geq 1$ such that the property holds for $L-1$. Consider $\widetilde Z_L(z_1,\dots,z_L)$ and $s_{1^L}^C(z_1,\dots,z_L|{\bf v})$ as Laurent polynomials
 in the variable $z_L$. It is easy to see that 
 \begin{eqnarray*}
  \widetilde Z_L(z_1,\dots,z_L) & = & \llangle \wt W | \left(z_1+ \frac{1}{z_1}+ \epsilon + \delta \right) \dots \left(z_L+ \frac{1}{z_L}+ \epsilon + \delta \right)|\wt V\rrangle \\
  & \displaystyle\underset{z_L \to\infty}{\sim} &  z_L \, \llangle\wt W | \left(z_1+ \frac{1}{z_1}+ \epsilon + \delta \right) \dots \left(z_{L-1}+ \frac{1}{z_{L-1}}+ \epsilon + \delta \right)|\wt V\rrangle \\
  & \displaystyle\underset{z_L \to\infty}{\sim} &  z_L\, \widetilde Z_{L-1}(z_1,\dots,z_{L-1}).
 \end{eqnarray*}
\end{itemize}
 We have also $\widetilde Z_L(z_1,\dots,z_L) \displaystyle\underset{z_L \to 0}{\sim}\widetilde Z_{L-1}(z_1,\dots,z_{L-1})/z_L$.
 Using the determinant formula for $s_{1^L}^C(z_1,\dots,z_L|{\bf v})$, we can develop the determinants in the numerator and denominator
 with respect to the last column to show that
 $s_{1^L}^C(z_1,\dots,z_L|{\bf v}) \displaystyle\underset{z_L \to\infty}{\sim} z_L s_{1^{L-1}}^C(z_1,\dots,z_{L-1}|{\bf v})$ and 
 $s_{1^L}^C(z_1,\dots,z_L|{\bf v}) \displaystyle\underset{z_L \to 0}{\sim}  s_{1^{L-1}}^C(z_1,\dots,z_{L-1}|{\bf v})/z_L$. Using
 the induction hypothesis we deduce that
 \begin{equation*}
  s_{1^L}^C(z_1,\dots,z_L|{\bf v}) \displaystyle\underset{z_L \to\infty}{\sim}\widetilde Z_L(z_1,\dots,z_L) \quad \text{ and } \quad 
  s_{1^L}^C(z_1,\dots,z_L|{\bf v}) \displaystyle\underset{z_L \to 0}{\sim}\widetilde Z_L(z_1,\dots,z_L).
 \end{equation*}
 Hence $\widetilde Z_L(z_1,\dots,z_L)$ and $s_{1^L}^C(z_1,\dots,z_L|{\bf v})$ are Laurent polynomials in the variable $z_L$ of degree $1$ in $z_L$ and $1/z_L$ 
 and they have the same coefficients in front of $z_L$ and $1/z_L$. So we can write
 \begin{equation*}
 \widetilde Z_L(z_1,\dots,z_L)- s_{1^L}^C(z_1,\dots,z_L|{\bf v})=f(z_1,\dots,z_{L-1}),
 \end{equation*}
with $f$ independent of $z_L$. But $\widetilde Z_L(z_1,\dots,z_L)$ and $s_{1^L}^C(z_1,\dots,z_L|{\bf v})$ are symmetric under the permutation of the $z_i$'s,
so $f$ cannot depend of the $z_i$'s and hence is constant. We compute this constant by evaluating the previous expression at the particular
point where all the $z_i$'s are equal to $1$. On one hand 
the value of $s_{1^L}^C(1,\dots,1|{\bf v})$ can be computed using \eqref{eq:ex} and \eqref{eq.dimC}. 
On the other hand the value of $\widetilde Z_L(1,\dots,1)$ is equal to the value of the usual continuous time TASEP normalization factor,
which matches with the previous value (see \eqref{eq:egaliteZL}). Hence we have
$\widetilde Z_L(1,\dots,1)=s_{1^L}^C(1,\dots,1|{\bf v})$, so $f=0$. This concludes the proof.

\section{Expansion of shifted Schur polynomials of type C\label{app.schur}}
We show that the type C shifted Schur polynomial \eqref{schurCshifte}  can be expanded on the regular type C Schur polynomials  as follows
 \begin{equation}\label{eq:expansionSchur}
s^C_{1^L}(z_1,\dots,z_L|\va)=\sum_{n=1}^{L+1} \frac{a^n-b^n}{a-b}\, s^C_{1^{L+1-n}}(z_1,\dots,z_L),
\end{equation}
 where $\va=(-a,-b,0,\dots,0)$.
 \paragraph{Proof.}
 Starting from
 \begin{equation*}
s^C_{1^L}(z_1,\dots,z_L|\va)= \frac{\det \left((z_j|\va)^{L+2-i}-(1/z_j|\va)^{L+2-i} \right)_{i,j}}{\det \left(z_j^{L+1-i}-(1/z_j)^{L+1-i} \right)_{i,j}}
\end{equation*}
the general idea is to use the multi-linearity of the determinant in the numerator to expand it and recognize non shifted Schur polynomials of type C.
First we rewrite
\begin{eqnarray}
 (z_j|\va)^{L+2-i}-(1/z_j|\va)^{L+2-i} & = & z_j^{L+2-i}-\frac{1}{z_j^{L+2-i}}+(a+b)\left( z_j^{L+1-i}-\frac{1}{z_j^{L+1-i}} \right) \label{developpement} \\
 & & + ab\left(z_j^{L-i}-\frac{1}{z_j^{L-i}}\right). \nonumber
\end{eqnarray}
Define the matrix $M(z_1,\dots,z_L)=\left((z_j|\va)^{L+2-i}-(1/z_j|\va)^{L+2-i} \right)_{i,j}$. Let $R_i(z_1,\dots,z_L)$ be the 
$i$-th row of $M(z_1,\dots,z_L)$. The expansion \eqref{developpement}, we can rewritten as
\begin{equation} \label{dev_ligne}
 R_i(z_1,\dots,z_L)=P_{i-1}(z_1,\dots,z_L)+(a+b)P_i(z_1,\dots,z_L)+abP_{i+1}(z_1,\dots,z_L),
\end{equation}
with $P_k(z_1,\dots,z_L)$ the row vector $(z_j^{L+1-k}-1/z_j^{L+1-k})_j$. Remark that $P_{L+1}(z_1,\dots,z_L)=0$.
We develop $\det \left(M(z_1,\dots,z_L) \right)$ using the expansion \eqref{dev_ligne} and the multi-linearity of the determinant. 
{Since we have only $L+1$ different row vectors $P_i$, antisymmetry of the determinant imposes that we need 
to choose $L$ distinct row vectors $P_i$ among $P_0,\dots,P_L$ in order to get a non vanishing determinant 
(hence there is just one $P_i$ missing in each non vanishing determinant). Then, the expansion 
of $\det(M(z_1,\dots,z_L))$ reads
\begin{equation*}
 \det(M(z_1,\dots,z_L))= \sum_{n=1}^{L+1}C_n\, u_n(z_1,\dots,z_L),
\end{equation*}
where
\begin{equation} \label{determinant_type}
u_{L+1-k}(z_1,\dots,z_L) = \det \left( \begin{array}{c}
              P_0(z_1,\dots,z_L) \\
              \vdots \\
              P_{k-1}(z_1,\dots,z_L) \\
              P_{k+1}(z_1,\dots,z_L) \\
              \vdots \\
              P_L(z_1,\dots,z_L)
             \end{array} \right), \qquad \text{with} \qquad k\in \{0,..,L\}
\end{equation}
and $C_n$ are some coefficients to be determined.}

Let us first compute $C_1$ (which corresponds to the term with missing $P_L$ in \eqref{determinant_type}). There is only one possibility to get
the row vector $P_0$: it comes necessarily from the row $R_1$. Then $P_1$ appears in the rows $R_1$ and $R_2$, but 
since we have already chosen $P_0$
for the row $R_1$, $P_1$ has to be taken in row $R_2$. One shows by induction that for all $i\in \{0,\dots L-1\}$,
$P_i$ has to be taken in row $R_{i+1}$. It is summarized on the following diagram.
\begin{equation}
 \begin{array}{cccccccc}
  R_1 & \quad & \bf{P_0} & P_1 & P_2 &  & \quad &    \\
  R_2 &       &     & \bf{P_1} & P_2 & P_3 &    &    \\
  \vdots &    &     &     & \ddots & \ddots & \ddots &  \\
  R_{L-1}&    &     &     &        & \bf{P_{L-2}} & P_{L-1} & \xout{P_L}  \\
  R_L    &    &     &     &        &         & \bf{P_{L-1}} & \xout{P_L} 
 \end{array}\label{eq:PL}
\end{equation}
{Each line in \eqref{eq:PL} corresponds to the expansion \eqref{dev_ligne} of a row vector $R_i$. The coefficients have been omitted, but can be easily recovered from \eqref{dev_ligne}.}
The bold $P_i$'s in the diagram are the ones chosen in the expansion to obtain the term $u_1(z_1,\dots,z_L)$. The shaded $P_i$'s are the ones
that do not appear in $u_1(z_1,\dots,z_L)$. 
{In the case considered here, all the chosen $P_i$'s correspond to the first term in \eqref{dev_ligne}} 
and hence come with a factor $1$.
We deduce that $C_1=1$.

The computation of $C_2$ {(with $P_{L-1}$ missing)} looks similar. There is again only one {choice for the $P_i$'s} to get $u_2(z_1,\dots,z_L)$. It
is given by
\begin{equation}
 \begin{array}{ccccccccc}
  R_1 & \quad & \bf{P_0} & P_1 & P_2 &  & \quad &   & \\
  R_2 &       &     & \bf{P_1} & P_2 & P_3 &    &   & \\
  \vdots &    &     &     & \ddots & \ddots & \ddots & & \\
  R_{L-2}&    &     &     &        & \bf{P_{L-3}} & P_{L-2} & \xout{P_{L-1}} & \\
  R_{L-1}&    &     &     &    &   & \bf{P_{L-2}} & \xout{P_{L-1}} & P_L  \\
  R_L    &    &     &     &    &   &         & \xout{P_{L-1}} & \bf{P_L} 
 \end{array}
\end{equation}
The chosen $P_0,\dots,P_{L-2}$ {correspond to the first term in \eqref{dev_ligne}} and hence come with a factor $1$, but the chosen $P_L$ {corresponds to the second term in \eqref{dev_ligne}} and thus
comes with a factor $a+b$. We deduce that $C_2=a+b$.

To compute $C_n$ {(with $P_{L+1-n}$ missing)}, we draw the diagram 
\begin{equation}
 \begin{array}{cccccccccccc}
  R_1 & \quad & \bf{P_0} & P_1 & P_2 &  & \quad &   &  & & & \\
  R_2 &       &     & \bf{P_1} & P_2 & P_3 &    &   &  & & & \\
  \vdots &    &     &     & \ddots & \ddots & \ddots & & & & & \\
  R_{L-n}&    &     &     &        & \bf{P_{L-1-n}} & P_{L-n} & \xout{P_{L+1-n}} & & & & \\
  R_{L+1-n}&    &     &     &    &   & \bf{P_{L-n}} & \xout{P_{L+1-n}} & P_{L+2-n}& & & \\
  R_{L+2-n}&    &     &     &    &   &         & \xout{P_{L+1-n}} & P_{L+2-n}& P_{L+3-n} & & \\
  R_{L+3-n}&    &     &     &        &        &   &     & P_{L+2-n}  & P_{L+3-n} & P_{L+4-n} & \\
  \vdots &    &     &     &        &        &   &     &  & \ddots & \ddots & \ddots
 \end{array}
\end{equation}
{In opposition to the previous cases, there is some freedom in this diagram. Indeed, although the first rows $R_1,\dots,R_{L+1-n}$ uniquely fix which $P_i$'s are chosen (written in bold), the
 row $R_{L+2-n}$ corresponds to two choices:
\begin{itemize}
\item we choose $P_{L+2-n}$ (which comes with the factor $a+b$). Then, the remaining rows just correspond to 
  the computation of $C_{n-1}$.
\item we choose $P_{L+3-n}$ (which comes with the factor $ab$). In this case, we have to choose $P_{L+2-n}$ in the row
$R_{L+3-n}$ (which comes with the factor $1$). Then, the remaining rows correspond to the computation of $C_{n-2}$ (except that the order 
of $P_{L+2-n}$ and $P_{L+3-n}$ is inverted, hence a factor (-1) by antisymmetry of the determinant). 
\end{itemize}
Altogether, we get a recursion relation for the coefficients $C_n$:}
\begin{equation} \label{recursion_C}
 C_n=(a+b)\,C_{n-1}-ab\,C_{n-2}.
\end{equation}
It is straightforward to check that $C_n=\frac{a^n-b^n}{a-b}$ verifies \eqref{recursion_C}, $C_1=1$ and $C_2=a+b$. 
We complete the proof of \eqref{eq:expansionSchur} by remarking that
$u_n(z_1,\dots,z_L)/\det \left(z_j^{L+1-i}-(1/z_j)^{L+1-i} \right)_{i,j} =s^C_{1^{L+1-n}}(z_1,\dots,z_L)$.

\end{document}